\documentclass{jfm}

\usepackage{graphicx}
\usepackage{natbib}
\usepackage{amssymb}
\usepackage{upmath}
\usepackage{xcolor}
\usepackage{wasysym}

\ifCUPmtlplainloaded \else
  \checkfont{eurm10}
  \iffontfound
    \IfFileExists{upmath.sty}
      {\typeout{^^JFound AMS Euler Roman fonts on the system,
                   using the 'upmath' package.^^J}%
       \usepackage{upmath}}
      {\typeout{^^JFound AMS Euler Roman fonts on the system, but you
                   dont seem to have the}%
       \typeout{'upmath' package installed. JFM.cls can take advantage
                 of these fonts,^^Jif you use 'upmath' package.^^J}%
      }
  \else
  \fi
\fi

\ifCUPmtlplainloaded \else
  \checkfont{msam10}
  \iffontfound
    \IfFileExists{amssymb.sty}
      {\typeout{^^JFound AMS Symbol fonts on the system, using the
                'amssymb' package.^^J}%
       \usepackage{amssymb}%

      }{}
  \fi
\fi

\ifCUPmtlplainloaded \else
  \IfFileExists{amsbsy.sty}
    {\typeout{^^JFound the 'amsbsy' package on the system, using it.^^J}%
     \usepackage{amsbsy}}
    {\providecommand\boldsymbol[1]{\mbox{\boldmath $##1$}}}
\fi

\newsavebox{\astrutbox}
\sbox{\astrutbox}{\rule[-5pt]{0pt}{20pt}}

\title[Intermittency in the relative separations of tracers and of heavy
  particles]{Intermittency in the relative
  separations of tracers and of heavy particles in turbulent flows}

\author[L. Biferale, A. S. Lanotte, R. Scatamacchia, and F. Toschi]
       {L. Biferale$^1$, A. S. Lanotte$^2$, R. Scatamacchia$^{1,3}$,\ns
         \break and F. Toschi$^{3,4,5}$}

\affiliation{$^1$Department of Physics and INFN, Univ. of Tor
  Vergata,\\ Via della Ricerca Scientifica 1, 00133 Rome,
  Italy\\
[\affilskip] $^2$
  CNR-ISAC and INFN- Sez. Lecce, Str. Prov. Lecce-Monteroni, 73100 Lecce,
  Italy\\
[\affilskip] $^3$Department of Physics, Eindhoven Univ.
  of Technology, 5600 MB Eindhoven, The Netherlands\\
[\affilskip]$^4$Department of Mathematics and Computer
  Science, Eindhoven Univ. of Technology,\\ 5600 MB Eindhoven, The
  Netherlands\\
[\affilskip]$^5$ CNR-IAC, Via dei Taurini 19, 00185 Rome, Italy}

\pubyear{}
\volume{}
\pagerange{}
\date{for publication in this finale form (postprint version) on J. Fluid Mech. ${\bf 757}$, $550$ $(2014)$}
\begin{document}

\maketitle

\begin{abstract}
Results from Direct Numerical Simulations of
particle relative dispersion in three dimensional homogeneous and
isotropic turbulence at Reynolds number $Re_\lambda \sim 300$ are
presented. We study point-like passive tracers and heavy particles, at
Stokes number $St=0.6,\,1$ and $5$. Particles are emitted from
localised sources, in bunches of thousands, periodically in time,
allowing to reach an unprecedented statistical accuracy, with a total
number of events for two-point observables of the order of
$10^{11}$.\\ The right tail of the probability
  density function for tracers develops a clear deviation from
  Richardson's self-similar prediction, pointing to the intermittent
  nature of the dispersion process. In our numerical experiment, such
  deviations are manifest once the probability to measure an event
  becomes of the order of -or rarer than- one part over one million,
  hence the crucial importance of a large
  dataset.\\The role of finite-Reynolds effects and
  the related fluctuations when pair separations cross the boundary
  between viscous and inertial range scales are discussed. An
  asymptotic prediction based on the multifractal theory for inertial
  range intermittency and valid for large Reynolds numbers is found to
  agree with the data better than the Richardson theory. The agreement
  is improved when considering heavy particles, whose inertia filters
  out viscous scale fluctuations.
%
By using the exit-time statistics we also show that events associated
to pairs experiencing unusually slow inertial range separations have a
non self-similar probability distribution function.
\end{abstract}


\section{Introduction}
Dispersion of particles in stochastic and turbulent flows is a key
fundamental problem \citep{MoninYaglom} with applications in a huge
number of disciplines going from atmospheric and ocean sciences
\citep{Bennet1984,LaMaRi2008,Ollitraut,PZ,Lacasce2010}, to
environmental sciences \citep{csanady73}, chemical engineering and
astrophysics \citep{Baldyga,Lepreti}. At high Reynolds numbers,
molecular diffusion makes a negligible contribution to spatial
transport, and so turbulence dominates not only the transport of
momentum, but also that of temperature, humidity, salinity and of any
chemical species or concentration field. Mixing can be approached from
an Eulerian point of view, studying the spatial and temporal evolution
of a concentration field \citep{dimotakis05}, and also by using a
Lagrangian approach in terms of the relative dispersion of pairs of
particles \citep{rev_sawford,FGV,rev_collins}. Notwithstanding the
enormous literature on the topic, a stochastic model for particle
trajectories in turbulent flows whose basic assumptions are fully
justified is yet to come \citep{Th90,BS94,Kurb97,BY04b,P08}. The
modeling of pair dispersion for tracers was pioneered in
\citet{Rich26}, where a {\it locality assumption} was introduced, see
also \citet{benzi11} for a recent historical review. In a modern
language, Richardson's approach is built up in analogy with diffusion,
replacing molecular fluctuations with turbulent fluctuations, acting
differently at different scales. Hence, in a turbulent flow,
diffusivity is enhanced because the instantaneous separation rate
depends on the local turbulent conditions encountered by pairs along
their path:
\begin{equation}
\frac{d \langle r^2 \rangle}{dt} = D(r)\,,\quad r_0 \ll r \ll L\,, 
\label{eq:1}
\end{equation}
where $r(t)$ is the amplitude of the separation vector between the two
particles, ${\bf r}(t) = {\bf X}_1(t) - {\bf X}_2(t)$, and $D(r)$ is a
scalar eddy-diffusivity. For the eddy-diffusivity approach to be
valid, separations $r$ have to be chosen larger than the initial ones,
$r_0$, and smaller than the integral scale of the flow, $L$. Moreover,
time lags have to be large enough, so that the memory of the initial
separation is lost.\\ In the light of Kolmogorov 1941 theory, (see
\citet{frisch}, the scalar eddy-diffusivity can be  modeled 
as follows:
\begin{equation}
D(r) \propto \tau(r,t) \,\langle (\delta_r u)^2\rangle \sim k_0 \epsilon^{1/3} r^{4/3}\,,
\label{eq:43}
\end{equation}
 where $\delta_r u$ is the Eulerian longitudinal velocity difference
 along the direction of particle separation ${\bf r}$, $\delta_r
 u({\bf r}(t),t) = \hat {\bf r} \cdot ({\bf u}({\bf X}_1(t),t) -{\bf
   u}({\bf X}_2(t),t))$, $k_0$ is a dimensionless constant, and
 $\epsilon$ is the rate of kinetic energy dissipation in the flow. In
 the above equation, $\tau(r,t)$ is the correlation time of the
 Lagrangian velocity differences at scale ${\bf r}$\,,
\begin{equation}
\tau(r,t) = \frac{2}{\langle [\delta_r u({\bf r}(t),t)]^2\rangle} \int_0^t \langle \delta_r {\bf u}({\bf r}(t),t) \cdot \delta_r {\bf u}({\bf r}(s),s) \rangle\, ds\,.
\label{eq:tau}
\end{equation}
 In the inertial range of scales, by dimensional considerations, we
 can write $\tau(r) \simeq \epsilon^{-1/3}r^{2/3}$, and $\langle
 (\delta_r u)^2\rangle \simeq (\epsilon\,r)^{2/3}$, from which the
 celebrated Richardson's $4/3$ law of eqn.~(\ref{eq:43}) follows. As a
 consequence, eqn.~(\ref{eq:1}) predicts a super-diffusive growth for
 the particle separation:
\begin{equation}
\label{eq:t3}
\langle r^2(t) \rangle \simeq \epsilon\,t^3\,,
\end{equation}
 and the dependence on the initial conditions is quickly forgotten. In
 fact, when released into a fluid flow, tracer pairs separate
 ballistically at short time lags, {\it \' a la } Batchelor
 \citep{Batch50}, keeping memory of their initial longitudinal
 velocity difference, $\langle r^2(t) \rangle \simeq (\langle r_0^2
 \rangle + C (\epsilon r_0)^{2/3} t^2$, up to time lags of the order
 of $t_B(r_0) \sim (r_0^2/\epsilon)^{1/3}$. Only later on,
 Richardson's super-diffusive regime follows.\\ Richardson's approach
 is exact if we assume that tracers disperse in a $\delta-$correlated
 in time velocity field. In such a case, the probability density
 function (PDF) of observing two tracers at separation $r$ at time
 $t$, $P(r,t|r_0 t_0)$, satisfies a Fokker-Planck diffusive equation
 with a space dependent diffusivity coefficient, $D(r)$
 \citep{Kraich66,FGV}:
\begin{equation}
\label{eq:Rich}
\frac{\partial P(r,t)}{\partial t} = \frac{1}{r^2}
\frac{\partial}{\partial r}\left[ r^2 D(r) \frac{\partial
    P(r,t)}{\partial r}\right]\,,
\end{equation}
where $D(r)$ is a function of the velocity
  correlation evaluated at the current separation, only.\\
\noindent
The Richardson equation (\ref{eq:Rich}) with initial condition $P(r,
t_0) \propto \delta(r-r_0)$ can be solved, see e.g.,
\citet{lundgren81,bennett}, and the solution has an asymptotic, large
time form (independent of the initial condition $r_0$ and $t_0$) of
the kind:
\begin{equation}
  \label{eq:pdfRich}
P(r,t) = A\,\frac{r^2}{ (k_0 \epsilon ^{1/3} t)^{9/2}}\, \exp\left[-
  \frac{9r^{2/3}}{4k_0\epsilon ^{1/3} t}\right] \,,
\end{equation} 
where $A$ is a normalization constant. The Richardson PDF is perfectly
self-similar, so that all positive moments behave according to the
dimensional law, $\langle r^p \rangle \propto
(\epsilon^{1/3}\,t)^{3p/2}$.\\ There are many reasons for which the
Richardson distribution cannot exactly describe the behaviour of
tracer pairs in real flows. The most important ones are: (i) the
nature of the temporal correlations in the fluid flow
\citep{FGV,Chaves03}; (ii) the non-Gaussian fluctuations of turbulent
velocities \citep{frisch}; (iii) the small-scale effects induced by
the dissipation sub-range, and (iv) the large-scale effects induced by
the flow correlation length. These last two are of course connected to
finite Reynolds-number effects.\\ It is worth noticing that formally
any diffusion coefficient of the form $D(r,t) \sim
r^{\alpha}\,t^{\beta}$, with $3\alpha+2\beta=4$, is compatible with
the $\sim t^3$ law, however different results would then be obtained
for the functional form of $P(r,t)$ \citep{MoninYaglom}.\\
\begin{figure}
\begin{center}
\includegraphics[width=10cm,draft=false]{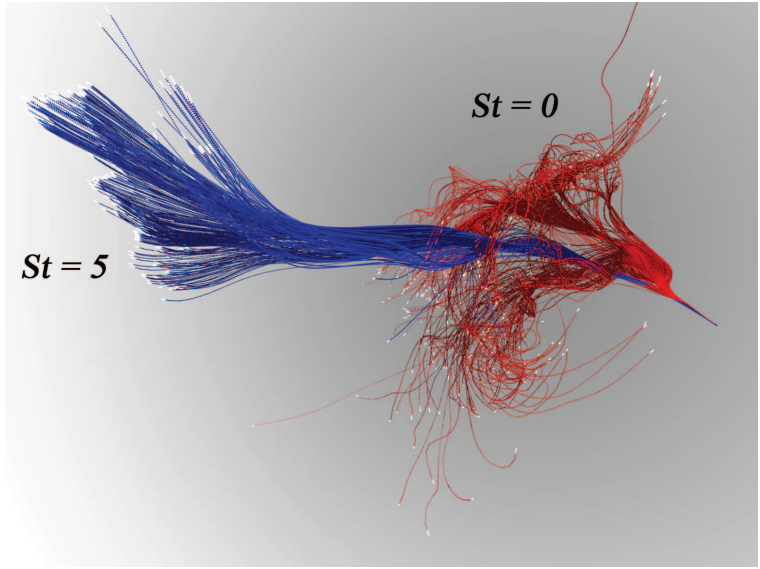}
\caption{(color online). An ensemble of tracer particles with $St = 0$
  (red) and heavy particles with $St = 5$ (blue), simultaneously
  emitted from a source of size $\sim \eta$. Trajectories are recorded
  from the emission time, up to the time $t = 75 \tau_\eta$ after the
  emission.}
\label{fig_two_stokes}
\end{center}
\end{figure}
Since Richardson's seminal work, pair dispersion has been addressed in
a large number of experimental and numerical studies, in the $2d$
inverse energy cascade \citep{JPT99,BoCe00_2d,BofSok2002_2d} and in
the direct enstrophy cascade (see e.g., \citet{Ju03}), as well as in
the $3d$ direct energy cascade \citep{Bif2005a,OM00}, in convective
turbulent flows \citep{Schumacher2008,Xia2013,MFLO14} and in synthetic
flows \citep{FV98,MV99,TD05,NN11}. In \citet{SYB05}, kinematic and
direct numerical simulations have also been used to compare forward
and backward relative dispersion in three dimensional turbulent
flows. Comprehensive reviews on the topic can be found in
\citet{rev_sawford},\citet{FGV} and \citet{rev_collins}.\\ Despite the
huge amount of theoretical, numerical and experimental works devoted to
this issue, it is fair to say that at the moment there is neither a
clear consensus in favour of the Richardson's approach, nor a clear
disproof. The main practical reason is due to the fact that the
predictions -if correct-, are applicable to tracer pairs whose
evolution has been for all times in the inertial range of scales:
 \begin{equation}
 \eta \ll r(t') \ll L \qquad \forall t' \in [t_0,t]\,,
\label{eq:IC} 
\end{equation}
where $\eta$ is the viscous scale of the turbulent flow. In other
words, we should record tracer dispersion at space and time scales
{\it unaffected} by viscous or integral scale effects. This is of
course a strong requirement which is particularly difficult to match
in any experimental or numerical test because of the natural
limitations in the accessible Reynolds number $Re_{\lambda}$, i.e., in
the scale separation range $Re_{\lambda} \propto (L/\eta)^{2/3}
$. Moreover the viscous scale itself $\eta$ and the stretching rate at
this scale are strongly fluctuating quantities in turbulent flows
\citep{frisch,Schu07,Yakhot2006,Bif2008}, causing further difficulties
when pair statistics must be limited to a pure inertial range
behaviour. It is worth noticing that a possible way out is to resort
to exit-time statistics \citep{vulpio,BofSok2002_2d,Bif2005a}, which
will be discussed in Section~\ref{sec:exit}. \\To avoid viscous
effects on the pair dispersion statistics, it is also common to study
pairs whose initial separation is well inside the inertial range,
$r(t_0) \gg \eta$, paying the price to be dominated for long times by
the initial condition and therefore mostly accessing the Batchelor
regime \citep{bodi_science,BBH12a}. Alternatively, numerical
simulations of particles evolving in stochastically generated velocity
fields are a useful tool to describe (possibly non Gaussian and non
self-similar) inertial range pair dispersion
\citep{Kurb97,BCCV99,MV99,TD05}. Note however that
  kinematic simulations might lead to a mean-square separation of the
  particle pairs with a power law different from the Richardson's law
  \citep{TD05}. \\ For the reasons (i)-(ii) listed above, it is well
possible that even in a infinite Reynolds number limit, the
Richardson's prediction may turn out to be wrong. Effects of time
correlations have been discussed by many authors
\citep{KBS87,Sok99,BBH12a,SBT12,Ey2013}, in connection to the problem
of the formally admissible {\it infinite propagation speed} present in
any diffusive approach {\it \`a la} Fokker-Planck
\citep{MW96,KOT09,IPZ13}, and also in relation to the
  possible non-Markovian nature of the Lagrangian position and
  velocity process \citep{Bec2014}. \\Summarising, it is extremely
important to clarify with high accuracy if the deviations from
Richardson's theory observed in laboratory experiments and numerical
simulations, at finite Reynolds numbers, are due to sub-leading
effects associated to the lack of scale-separation or not. In the
latter case, it means that they would survive even in the $Re
\rightarrow \infty$ limit. This is the aim of the research presented
in this paper. \\We use Direct Numerical Simulations of isotropic and
homogeneous three dimensional turbulence at $Re_\lambda \sim 300$,
seeded with an unprecedented number of particles (emitted from
localised sources in different locations inside the flow), in order to
increase the total number of pairs starting with an initially small
separation and to minimise local anisotropy and non-homogeneity.\\ We
present results for both tracers and point-like heavy particles,
without feedback on the flow as in the original problem attacked by
Richardson. When particles have inertia, new scenarios arise
\citep{FH08,JFM2010a}, because of the non homogeneous spatial
distribution \citep{JFM06} and the very intermittent nature of
relative velocity increments characterised by the presence of
quasi-singularities
\citep{Falknature,WM05,JFM2010b,JOPnoi,PanPadoan,SC2012}.\\ Not
surprisingly, and in the absence of a theory, empirical observations
are in this case even less stringent, also because of the need to
specify the initial distributions of both particle positions and
velocities. Two types of experiments can be done with inertial
particles. The first consists of studying relative dispersion as a
function of the distribution of initial separations only. In practice,
inertial particles are allowed to reach their stationary spatial and
velocity distributions inside a bounded volume (stationary
distribution on a fractal dynamical attractor in phase-space), after
which their dispersion properties are measured, conditioning on the
initial distance \citep{JFM2010a}. The second consists in directly
injecting inertial particles in the flow, with prescribed initial
velocity and separation distributions. The first protocol is more
relevant to study relative dispersion properties in connection with
spatial clustering, particularly effective at small Stokes numbers
(e.g., the spatial preferential concentration and trapping in coherent
structures in the flow), and in connection with caustics, strongly
modifying the relative velocities at large Stokes numbers
\citep{Abra75,JFM2010a,JFM2010b,PanPadoan}. The second protocol is
more relevant in geophysical and industrial applications, where
transient behaviours are crucial as in the case of volcanic eruptions,
leakages of contaminants, or pollutant emissions.\\ In this paper, we
are interested in the latter case, for which we designed the simplest
procedure of having inertial particles emitted in the same positions
and with the same velocities of the tracers (see
Figure~\ref{fig_two_stokes}). This choice turns out to be optimal to
better understand the statistics of tracers also, as it will become
clearer in the sequel.

The main results of the paper can be briefly anticipated.  First, we
quantify the importance of viscous-scale fluctuations for tracer
separations, showing that they easily affect the separation evolution
for time scales much larger than what predicted by the dimensional
Batchelor time, $t_B(r_0)$. This effect is so huge that, in current
experiments, it spoils any possibility to assess tracer scaling
properties in the {\it expected} inertial range of time scales. To
overcome this problem, we suggest that heavy pairs
  can be used as dispersing particles that are able to dynamically
  filter out viscous-time fluctuations. By measuring pair dispersion
  of heavy pairs at different degree of inertia, we are able to
  observe a much clearer convergence towards an inertial-range regime
  for the relative separation. We interpret this result as an
indication of the existence of infinite Reynolds-number corrections to
the Richardson's prediction, as expected on the basis of a
Eulerian-Lagrangian multifractal approach
\citep{Bo1993,BCCV99,BifPRL04}, yet never observed in real data. The
multifractal approach (MF) is also used to determine a set of specific
correlation moments involving powers of the pair separation distance
and relative velocity of tracer pairs, which are expected to be
statistically invariant along Lagrangian trajectories in the inertial
range \citep{FGV,FalkPRL2013}.\\ The paper is organised as follows.
In section~\ref{sec:numerics}, we present the details of our numerical
study. In section~\ref{sec:direct}, we introduce the relative
dispersion statistics for both tracers and heavy particles. In
sub-section~\ref{sec:direct-sub}, we discuss the effects due to finite
Reynolds numbers and those induced by the fluctuations of the viscous
scale. In sub-section \ref{sec:pdf}, we present the numerical data for
the probability density functions. In connection with the issue of
intermittent corrections, in sub-section~\ref{sec:direct-multi}, we
briefly review the multifractal prediction for the pair dispersion of
tracers developed by \citep{BCCV99}, and we test it against our
data. In section~\ref{sec:exit}, we discuss the results concerning the
exit-time statistics, probing pairs which separate very slowly.

\section{Numerical details}
\label{sec:numerics}

The fluid is described by the Navier-Stokes equations for the velocity
field ${\bf u}({\bf x}, t)$
\begin{equation}
\label{eq:ns}
\partial_t {\bf u} + {\bf u} \cdot {\bf \nabla} {\bf u} = - {\bf
  \nabla} p + \nu \nabla^2 {\bf u} + {\bf f}\,, \hspace{0.5cm} \nabla
\cdot {\bf u} =0\,.
\end{equation}
The statistically homogeneous and isotropic external forcing ${\bf f}$
injects energy in the first low-wavenumber shells, by keeping constant
in time their spectral content \citep{Chen1993}. The kinematic
viscosity $\nu$ is chosen such that the Kolmogorov length scale is
$\eta \simeq \delta x$, where $\delta x$ is the grid spacing, so that
a good resolution of the small-scale velocity dynamics is obtained.
\begin{table}
  \begin{center}
\def~{\hphantom{0}}
  \begin{tabular}{ccccccccccccc}
      $Re_\lambda$ & $N^3$ & $\eta$ & $\Delta x$ & $\epsilon$ & $\nu$
    & $\tau_\eta$ & $T_E$ & $u_{rms}$ & $N_p$ & $N_{sou}$ &
    $N_{tot}$ & $T_{traj}$\\[3pt] 280 & $1024^3$ & 0.005
    & 0.006 & 0.81 & 0.00088 & 0.033 & $67$  & 1.7 &
    2$\times 10^3$ & 256 &4$\times 10^{11}$ & 160\\
  \end{tabular}
  \caption{Parameters of the numerical simulations: Taylor-scale based
    Reynolds number $Re_{\lambda}$, grid resolution $N^3$, Kolmogorov
    length scale $\eta$ in simulation units (SU), grid spacing $\Delta
    x$ (SU), mean energy dissipation $\epsilon$ (SU), kinematic
    viscosity $\nu$ (SU), Kolmogorov time-scale $\tau_\eta$ (SU),
    large-scale eddy turnover time $T_E$ (in units of $\tau_{\eta}$),
    root-mean-square velocity $u_{rms}$ (SU), $N_p$ number of
    trajectories of inertial particles emitted for each Stokes number
    $St$ from each local source and for each puff, $N_{sou}$ number of
    sources in the flow, $N_{tot}$ total number of particle pairs emitted
    in all simulations per Stokes number ($10$ runs with $256$ local
    sources, each emitting $80$ puffs), $T_{traj}$ maximal temporal
    length of particle trajectories (in units of $\tau_{\eta}$).}.
\label{tab:param}
  \end{center}
\end{table}
The numerical domain is cubic, with periodic boundary conditions in
the three space directions; a fully dealiased pseudo-spectral
algorithm with second-order Adam-Bashforth time-stepping is used. We
performed a series of Direct Numerical Simulations with resolution of
$1024^3$ grid points and Reynolds number at the Taylor scale
$Re_\lambda \simeq 300$. The flow is seeded with bunches of tracers
and heavy particles, emitted in different fluid locations to reduce
the large scale correlations, and local inhomogeneous/anisotropic
effects. Each bunch is emitted within a small region of space, of
Kolmogorov scale size, in puffs of $2 \times10^3$ particles each, for
tracers and heavy particles. In a single run, there are $256$ of such
point sources, releasing about $80$ puffs with a frequency comparable
with the inverse of the Kolmogorov time. We collected statistics over
$10$ different runs. As a result, we follow a total amount of $4
\times 10^{11}$ pairs.\\ The heavy particles are assumed to be of
size much smaller than the Kolmogorov scale of the flow and with a
negligible Reynolds number relative to the particle size. In this
limit, their equations of motion take the simple form:
\begin{equation}
\label{eq:eq_part}
\dot {\bf X}(t) = {\bf V}(t)\,,   \hspace{0.5cm} \dot {\bf V}(t) = - \frac{1}{\tau_s}[{\bf V}(t) - {\bf u}({\bf X}, t)]\,,
\end{equation}
where the dots denote time derivatives. The particle position and
velocity vectors are ${\bf X}(t)$ and ${\bf V}(t)$, respectively;
${\bf u}({\bf X}, t)$ is the Eulerian fluid velocity evaluated at the
particle position. The particle response time is $\tau_s$. The flow
Kolmogorov time scale, appearing in the definition of the Stokes
number, $St = \tau_s / \tau_\eta$, is $\tau_\eta =(\nu /
\epsilon)^{1/2}$. Particle-particle interactions and the feedback of
the particles onto the flow are here neglected. In this work, we show
results for the following set of Stokes numbers: $St = 0.0, 0.6, 1.0$
and $5.0$. Additional details of the runs can be found in Table
\ref{tab:param}.

\section{Relative separation statistics}
\label{sec:direct}
In Figure \ref{fig_two_stokes}, we compare the time evolution up to a
time of the order of the large scale eddy turn over time, $T_E$, of a
bunch of tracers and a bunch of heavy particles with $St=5$, both emitted
in a region of strong shear.\\
\begin{figure}
\begin{center}
\includegraphics[width=10cm,draft=false]{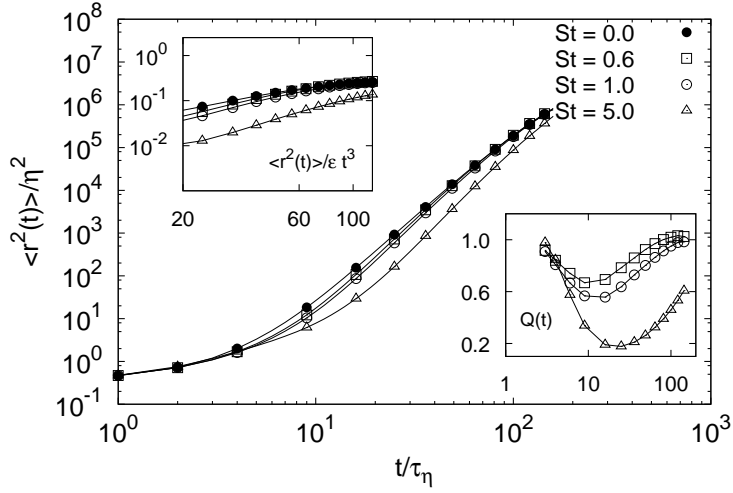}
 \caption{(Main body) Log-log plot of the mean-square separation
   versus time for particle pairs at changing inertia. (Upper inset)
   Log-log plot of the same curves, compensated with the Richardson's
   inertial range behavior. (Lower inset) The ratio $Q(t)$, in log-lin
   scale, of the mean-square separation of heavy particle pairs,
   normalised with the curve for tracer pairs.}
  \label{fig_r2_compared}
\end{center}
\end{figure}
At a time lag roughly equal to $t = 10 \tau_\eta$ after the emission,
it can be observed an abrupt transition in the particle dispersion of
tracers (red in the online version), occurring when most of the pairs
reaches a relative distance of the order of $10 \eta$. Later on, we
again notice the presence of many pairs with mutual separations much
larger than the mean one. The trajectories of the heavy particles
(blue in the online version) show a different evolution. After the
emission, they tend to remain at a mutual distance of the order of
$\eta$ for a very long time, thus dispersing much less. Because they
respond to fluid fluctuations with a time lag of the order of their
Stokes time, they tend to keep their initial velocity unchanged before
relaxing onto the underlying fluid velocities. As a result, inertial
particles behave as if their Batchelor time was much longer than that
of tracers (in our DNS the latter is small, $t_B(r_0,St=0) \sim
\tau_\eta$, because the source is strongly localised). Our main
observation here is that the larger the Stokes number, the longer is
the filtering time that heavy particles apply to the local stretching
properties of the carrying fluid: since inertia is moderate in the
present experiment, this fact will allow us to have a very effective
method to reduce effects from viscous scale fluctuations in the
particles' statistics and to better disentangle inertial range
properties in the pair dispersion evolution. \\ This is more
quantitatively understood in Figure~\ref{fig_r2_compared}, where the
second order moment of the relative separation for tracers, $St=0$,
and heavy particles, $St=5$, are plotted. Heavy particles tend to
separate less since they are unaffected by turbulent fluctuations up
to time and spatial scales large enough for their inertia to become
sub-dominant with respect to the underlying turbulent fluctuations. On
a dimensional ground, such a scale can be easily estimated to be of
the order of $r^*(St) \sim \eta St^{3/2}$ when $St > 1$
\citep{JFM2010a}. Let us notice that in this respect the choice of the
initially prescribed velocity distribution plays a key role, since
inertial particles are emitted with a velocity equal to that of the
underlying fluid. The quantity which is better suited to quantify this
observation is the ratio between relative separations:
$$ Q(t) = \frac{\langle r^2(t) \rangle_{St}}{\langle r^2(t)
  \rangle_{0}}\,.$$ In the present experiment we observe $Q(t)<1$ at
any scale and time, as shown in the lower inset of Figure
\ref{fig_r2_compared}. By prescribing an initial distribution equal to
the stationary PDF of heavy particle velocity increments, the trend
would have been the opposite, with inertial particles at short times
separating much faster than tracers, i.e. $Q(t)>1$ up to a separation
$r \simeq r^*(St)$ \citep{JFM2010b}. In the main body of Figure
\ref{fig_r2_compared}, we notice that for both tracers and heavy
particles, and for time lags large enough, the separation curves tend
toward a $t^3$ Richardson-like behaviour, but without showing any
clear scaling, as also measured by the compensated plot in the upper
inset of Figure~\ref{fig_r2_compared}.
\begin{figure}
\includegraphics[width=6.4cm,draft=false]{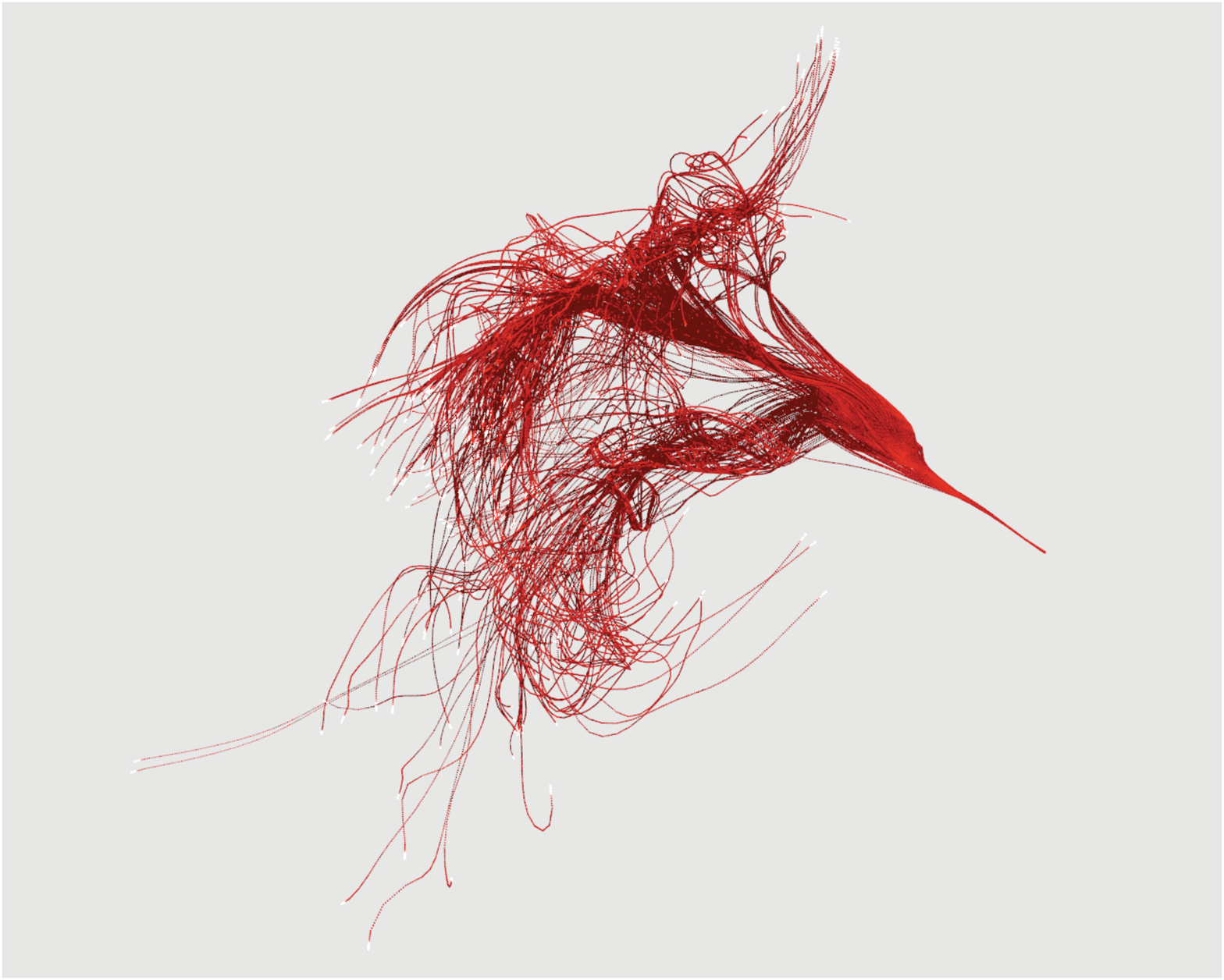}
\hspace{0.3cm}
\includegraphics[width=6.4cm,draft=false]{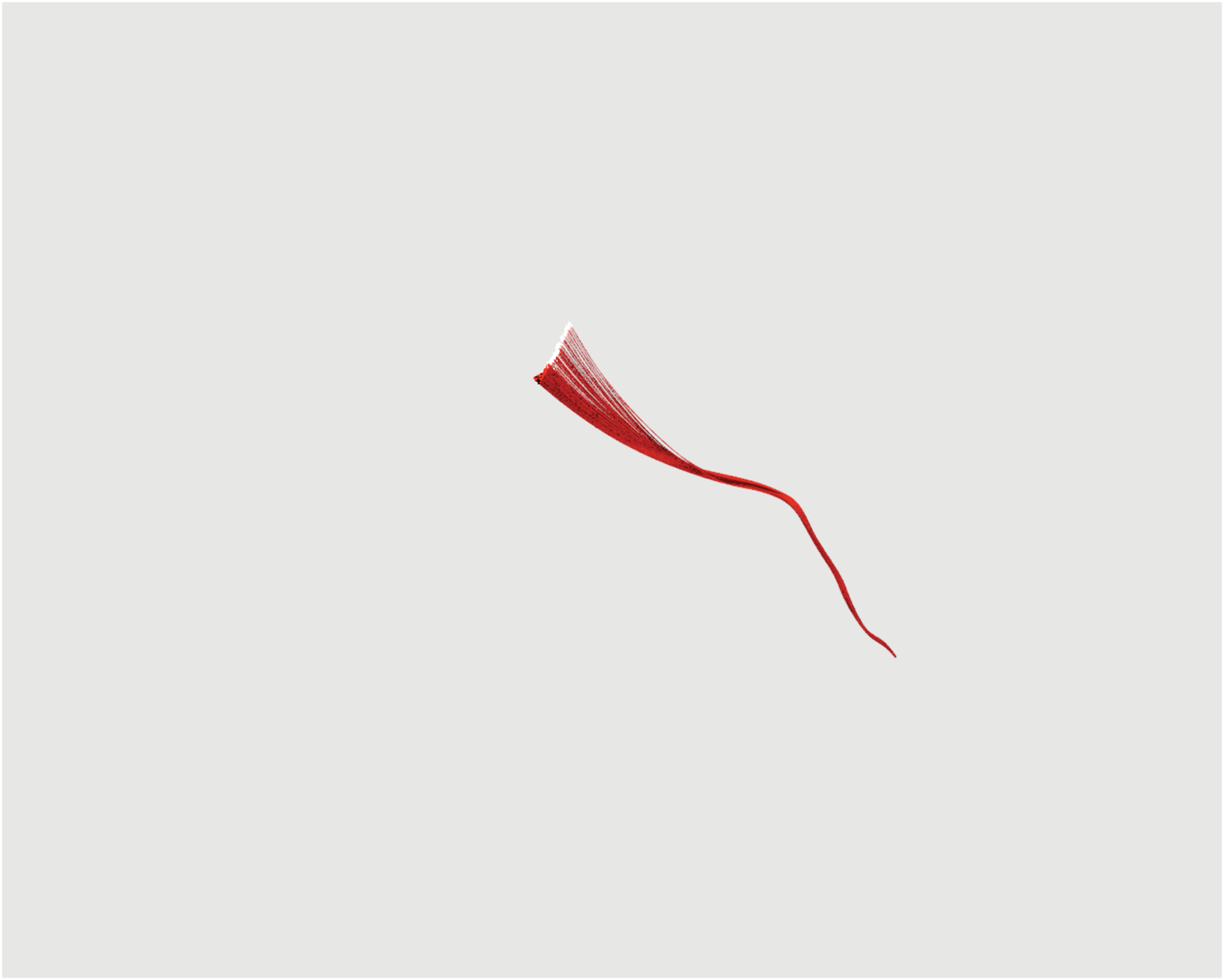}
\caption{(color online) Left panel: The same ensemble of tracers
  reported in Figure~\ref{fig_two_stokes}. Right panel: Simultaneous
  realisation of a tracer bunch, emitted from a different source,
  showing a much smaller dispersion. Both emissions are recorded up to
  the time $t = 75 \tau_{\eta}$.}
\label{fig_2_bunches}
\end{figure}

\subsection{Viscous effects}
\label{sec:direct-sub}

\begin{figure}
\begin{center}
\includegraphics[width=10cm,draft=false]{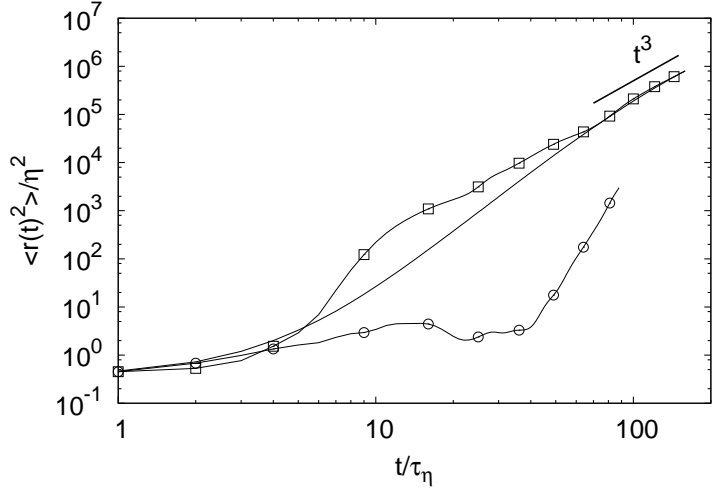}
 \caption{The mean-square separation behaviour for the two tracer
   emissions reported in Figure~\ref{fig_2_bunches}. Data from the
   left panel are represented by ($\square$); while data from the
   right panel are represented by ($\Circle$). The continuous curve is the mean square separation
   averaged over the whole statistical database.}
  \label{fig_2_bunches_r2}
\end{center}
\end{figure}

\begin{figure}
\begin{center}
\includegraphics[width=12cm,draft=false]{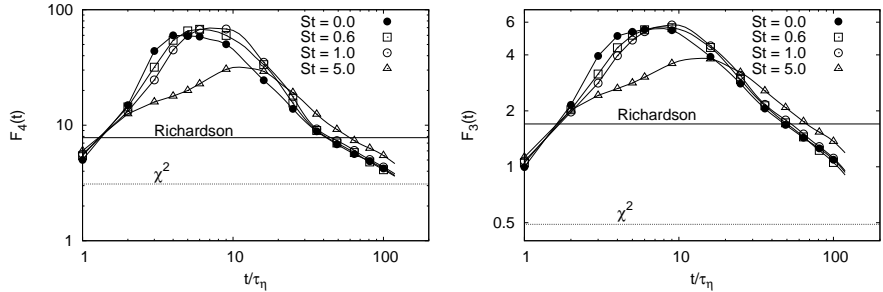}
\caption{Generalised flatness (left) and skewness (right) of the relative
  dispersion probability density function, for pairs of different
  inertia. Horizontal lines refer to the Richardson expectations for
  these observable, which are $7.81$ and $1.7$, respectively; also
  plotted are the expected values for a $\chi$-squared distribution with
  three degrees of freedom, which are $3.1$ for the flatness and
  $0.49$ for the skewness.}
\label{fig_flat}
\end{center}
\end{figure}

To better appreciate the importance of viscous effects on pair
dispersion, we show in Figure~\ref{fig_2_bunches}, the time evolution
of two different bunches of tracers emitted in different positions in
the flow. The initial size of the puffs is of the order of the viscous
scale. The bunch on the left is emitted in a region where the
stretching rate has a typical value of the order of its root mean
square, $\sim (\epsilon/\nu)^{-1/2}$, while the bunch on the right is
emitted in a region where the local stretching rate is unusually
small. As a result this second bunch separates with a much longer
delay with respect to the average behaviour.\\ In
Figure~\ref{fig_2_bunches_r2}, the mean squared separations measured
for pairs belonging to each of these two bunches are compared with the
pair separation averaged over the full statistics. The bunch emitted
in a region where the stretching rate assumes the typical value exits
the viscous region in a time lag of the order of $\tau_\eta$, and soon
approaches the inertial range behaviour compatible with $\sim
t^3$. Pairs belonging to the other bunch keep a small separation
$\langle r^2 \rangle^{1/2} \simeq \eta$ for a time lag up to $ \sim 50
\tau_{\eta}$, a time comparable to the integral time-scale $T_E$, and
never recover the $t^3$ scaling behaviour along the whole duration of
our simulation. The examples shown in Figure \ref{fig_2_bunches_r2}
are meant to represent the huge variations that affect pair separation
statistics for time lags of the order of $\tau_\eta$. These variations
are the hardest obstacle to assess pure {\it inertial} range
properties in any experimental or numerical set-up, in addition to the
challenge of following particle trajectories for a time lag long
enough -- and in a volume large enough.\\ Our idea here is to use
heavy particles as {\it smart} passive, but dynamical, objects to
filter out such huge viscous effects, without affecting the long-time
and large-scale physics. Indeed, heavy particles are less affected by
fluctuations of the local viscous scale, since they respond to the
fluid with their Stokes time (see Figure
\ref{fig_two_stokes}). Moreover, heavy pairs experience a less
fluctuating local stretching rate, as also measured by the
distribution of the finite-time Lyapunov exponents as a function of
the Stokes number \citep{BecPoF2006}. Finally, we recall that because
of the injection choice here adopted, caustics in the heavy particles
velocity distribution manifest themselves only at a later stage.
As for the heavy pairs dispersion, the different degrees of
fluctuations are better quantified in Figure~\ref{fig_flat},
where we show the ratio of the third and fourth order moments of the
separation distribution along the particle trajectories, normalised
to the second order one\,
\begin{equation}
F_3(t) = \frac{\langle r^3(t)\rangle}{\langle r^2(t)\rangle^{3/2}}; \qquad
F_4(t) = \frac{\langle r^4(t) \rangle}{\langle r^2(t) \rangle^2}\,.
\label{eq:flat_skew}
\end{equation} 
For convenience, we refer to $F_3(t)$ and $F_4(t)$ as
  generalised skewness and flatness, respectively. At the transition
between the viscous and the inertial range of scales, i.e., for $t
\simeq 10 \tau_\eta$, tracers and small Stokes particles possess a
generalised flatness $F_4(t)\sim 100$. This is the signature of an
extremely intermittent distribution (for a $\chi-$squared distribution
with three degrees of freedom, we would have $F_4=3.1$, while for the
Richardson distribution $F_4=7.81$). Moreover, it is evident that the
bump displayed by the generalised flatness around $10\tau_{\eta}$ is
influenced by the behaviour at shorter times and influences the behaviour at
much larger times. In other words, it is the quantitative counterpart
of the big variations shown in the two examples in
Figure~\ref{fig_2_bunches_r2}. \\ The filtering effect of the particle
Stokes time is the reason for which heavy pairs initially possess a
smaller flatness. This is particularly evident for the $St=5$ case,
which exhibits a smoother transition toward a scaling behaviour for $t
> 10-20 \tau_{\eta}$, supporting the idea the inertia helps to remove
viscous fluctuations from the physics of the inertial
range. At time lags $~20 \tau_{\eta}$, the
  generalised coefficients $F_3(t)$ and $F_4(t)$ for heavy particles
  with $St=5$ are larger than those of the tracer pairs. This is
  probably the result of the relaxation onto the tracer power-law
  behaviour, which happens only at a later stage. Finally, it is
clear from Figure \ref{fig_flat} that the data set at the moderate
inertia of $St=5$ is less affected by the strong viscous bump at $t
\sim 10 \tau_\eta$ and that therefore promises to be the best
candidate to test inertial range statistical properties.

\subsection{Probability density functions}
\label{sec:pdf}
\begin{figure}
\begin{center}
\includegraphics[width=6.3cm]{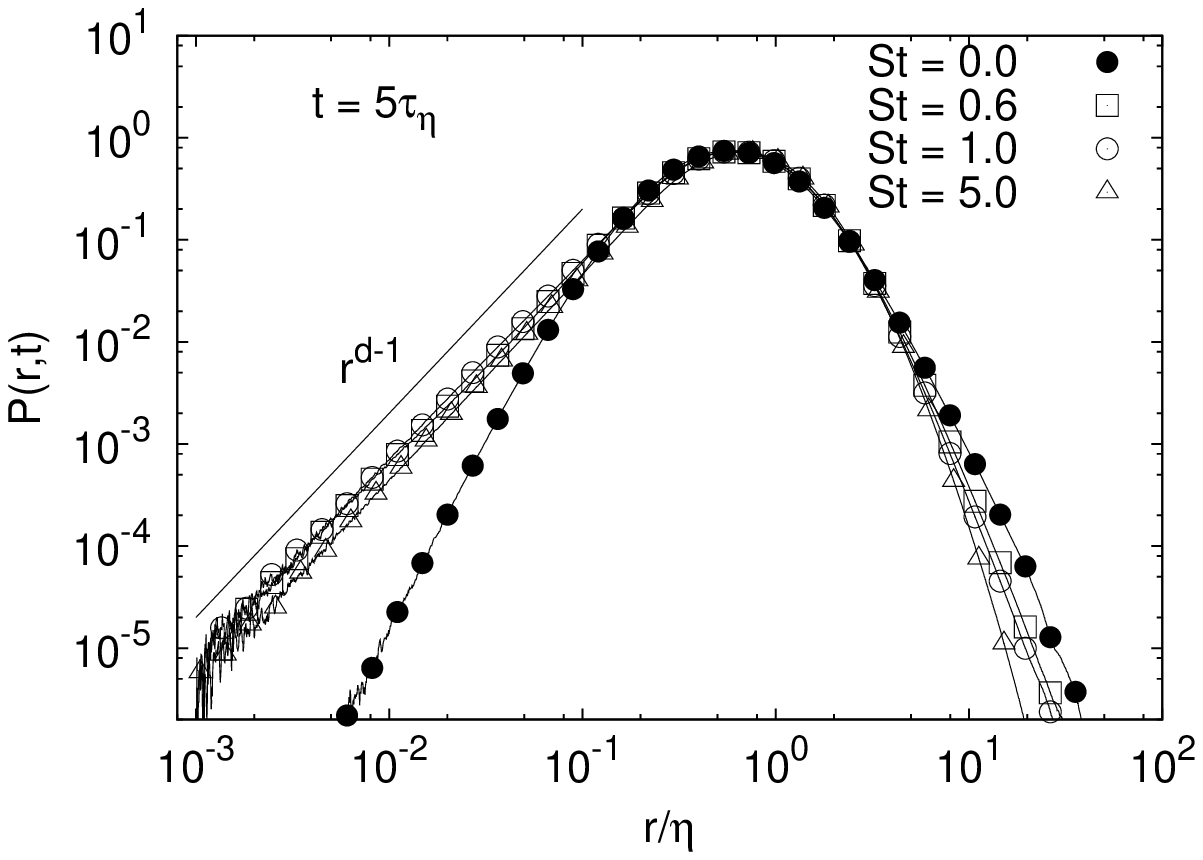}
\hspace{0.6cm}
\includegraphics[width=6.3cm]{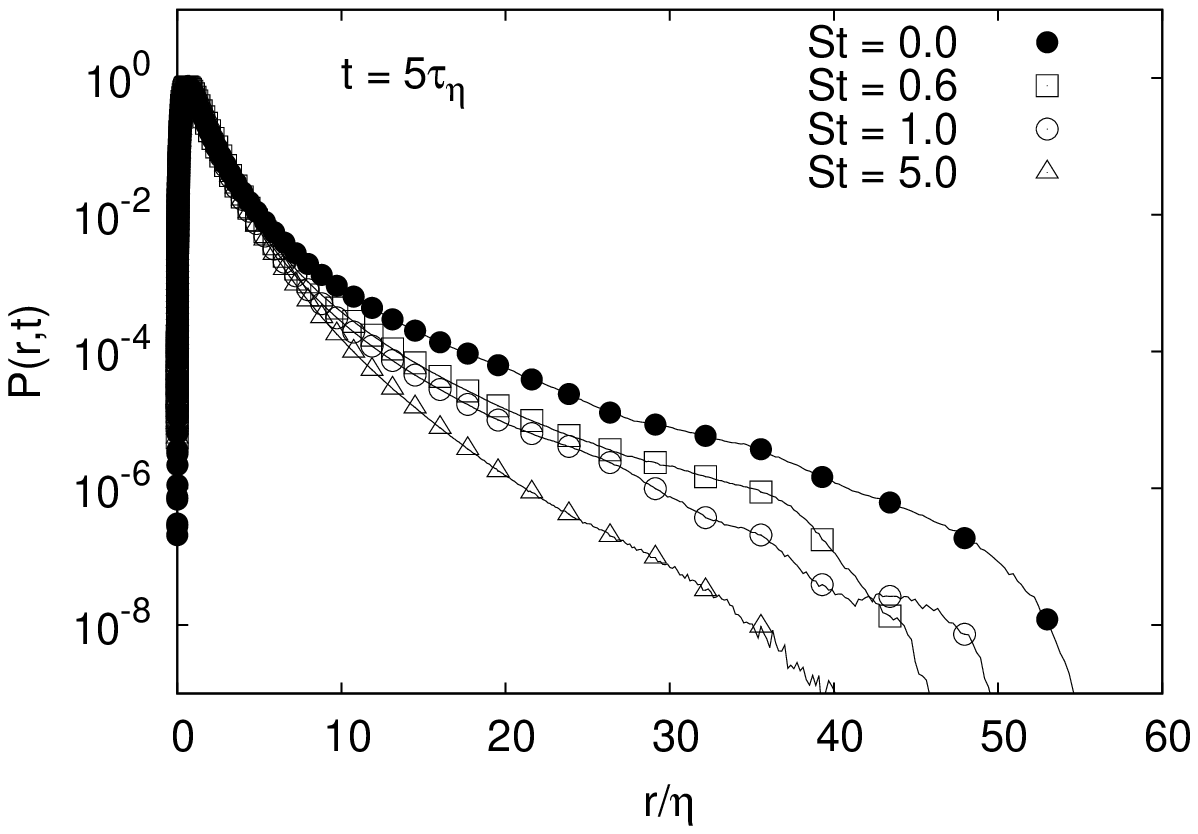}\
\includegraphics[width=6.3cm]{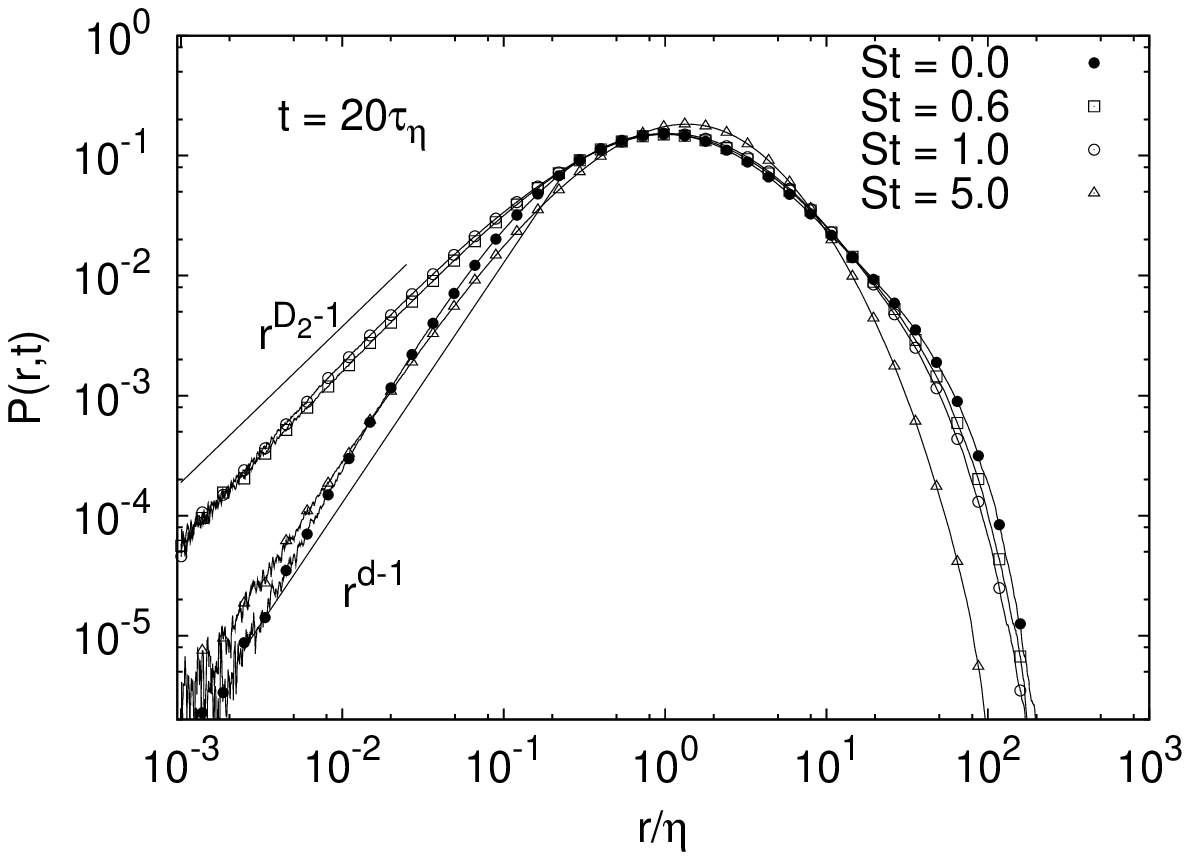}
\hspace{0.6cm}
\includegraphics[width=6.3cm]{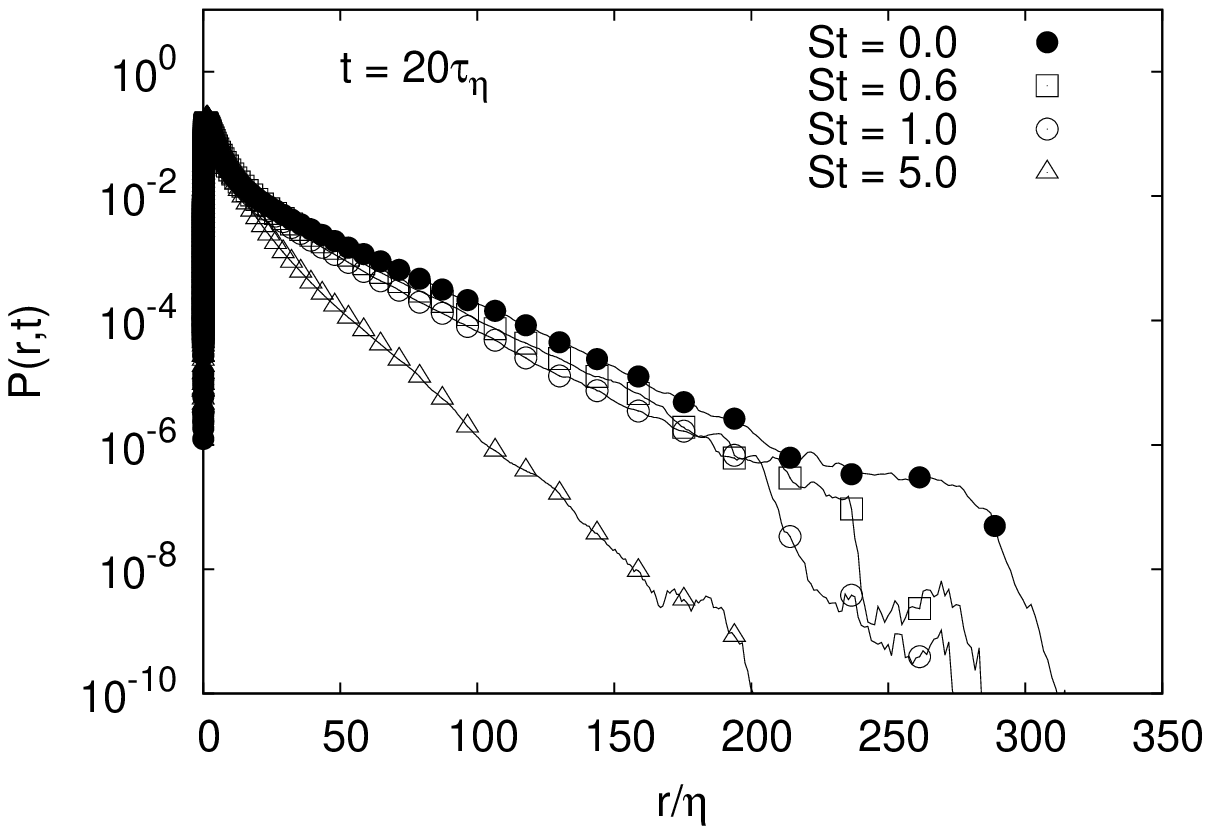}\
\includegraphics[width=6.3cm]{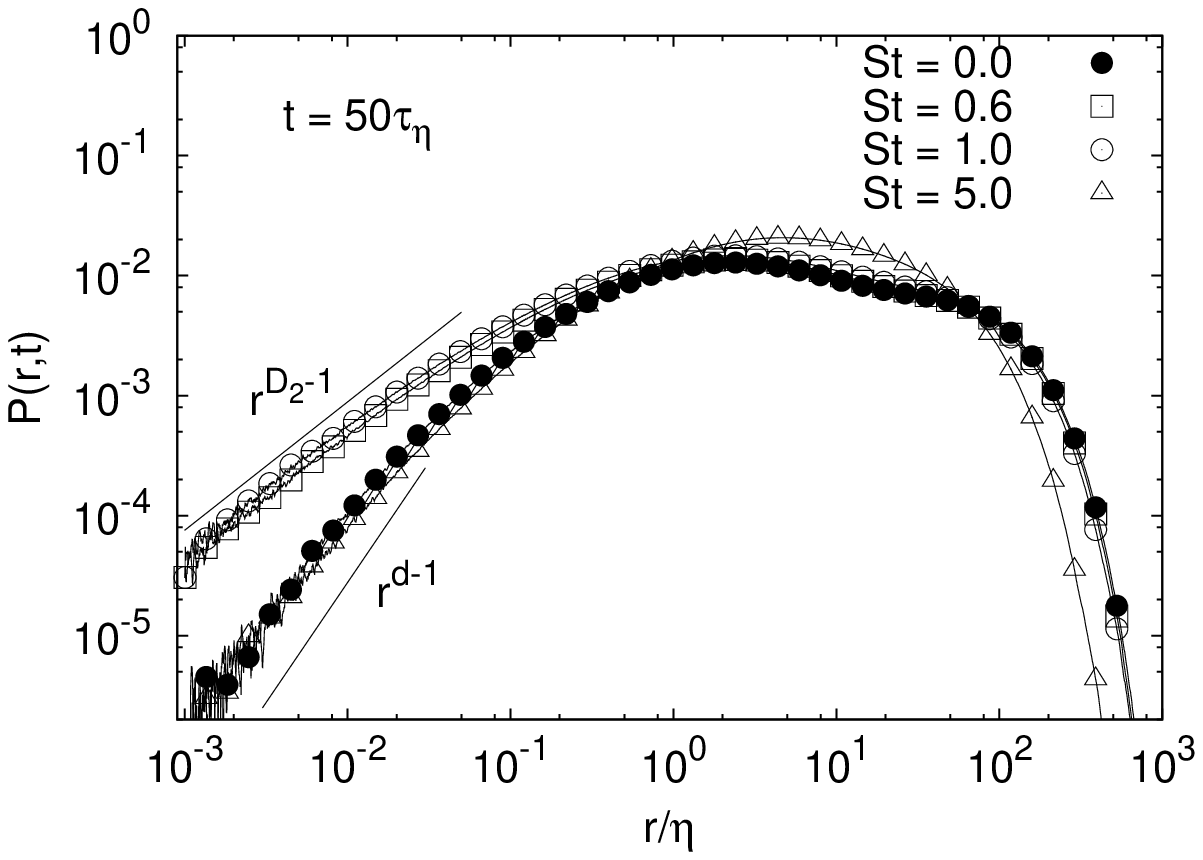}
\hspace{0.6cm}
\includegraphics[width=6.3cm]{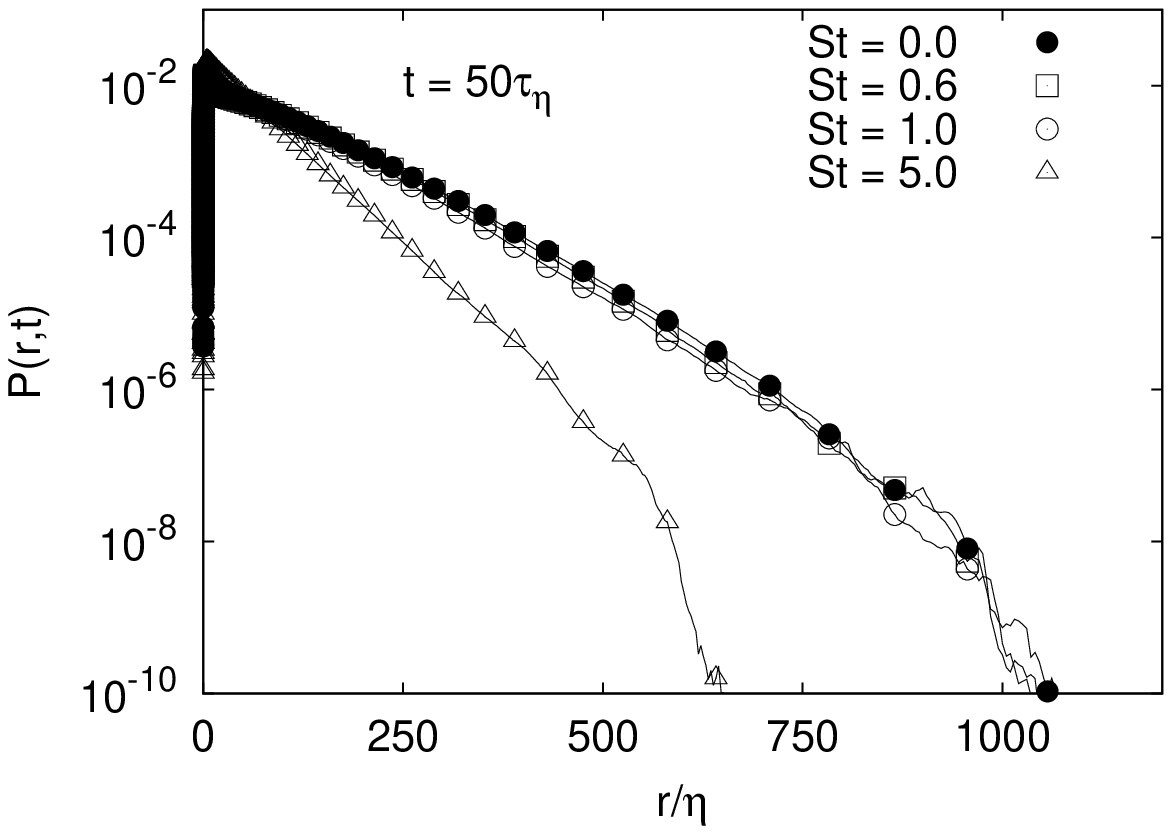}
\caption{(Left panels) Log-log plot of separation PDFs $P(r,t)$ for
  pairs with different inertia $St=0.0, 0.6, 1.0, 5.0$ highlighting
  the left tails behaviour. Plots refer to times $5 \tau_\eta$, $20
  \tau_\eta$ and $50 \tau_\eta$ after the emission. The power-law
  scaling $r^{d-1}$, with $d=3$, is plotted. The power-law scaling
  $r^{D_{2}-1}$ is also reported: note that for $St=0.6$ the
  correlation dimension $D_2(St)$ is $2.27\pm 0.03$, while for
  $St=1.0$ it is $D_2 (St) = 2.31 \pm 0.03$ \citep{JOPnoi}. The two
  power laws $r^{D_{2}-1}$ for $St=0.6$ and $St=1.0$ are
  indistinguishable in the scale of the plot, hence we reported the
  slope for $St=1.0$ only. (Right panels): Lin-log plot
  of the same separation PDFs, at the same time lags. For these PDFs
  pairs are selected with initial separation $r_0 \in [0.2:2] \eta$. }
  \label{fig_pdf_naked}
\end{center}
\end{figure}

The probability density functions $P(r,t)$ are plotted in
Figure~\ref{fig_pdf_naked}, for different values of inertia
$St=0,0.6,1$ and $5$, and at different time lags after the emission
from the source. In order to highlight the dynamics
  of those particles filling the left tail (i.e. separating less than
  the average), for Figure~\ref{fig_pdf_naked} we have selected pairs
  with initial separation $r(t_0) \in [0.2:2]\eta$ \citep{BBH12b}.
Consider first the behaviour at large separations. One clearly sees
the effect anticipated earlier. Heavy particles tends to separate
less. The effect becomes less and less visible with time, because
inertia is forgotten on a time scale roughly proportional to the
Stokes time. Pairs with $St=5$, however, accumulate a delay in
separation that is never recovered, even at large times $ t\sim 50
\tau_\eta$.\\ For the left tails, associated to pairs that do not
separate, the trend as a function of Stokes is the opposite. We recall
that from classical arguments \citep{FGV}, we expect that the fraction
of tracer pairs at a distance $r$ would behave as a power law
$r^{d-1}$, where $d = 3$ is the space dimension, see
eq.~(\ref{eq:pdfRich}). Similarly, for heavy pairs, we expect to
observe the scaling $P(r) \propto r^{D_{2}-1}$, where $D_2(St) \neq d$
measures the spatial correlation dimension. At short time lags, $t
\simeq 5 \tau_{\eta}$, the effect of inertia quickly appears and we
measure a higher probability to observe pairs at very small
separations: this happens because heavy pairs are
  less affected than tracers by intermittent events of anomalously
  slow separations, and hence rapidly populate the left tail of the
  distribution. As time goes on, $t \simeq 20 \tau_{\eta}$, we
observe that pairs with moderate inertia, namely $St=0.6$ and $St=1$,
clearly show the tendency to clusterise on a fractal set
\citep{BFF01,Bof2004,B2005,Chun2005,JFM06} characterised by the
spatial correlation dimension $D_2(St)<d$ \citep{JFM06,BecPRL2007},
where $d$ is the spatial dimension of the flow.
\begin{figure}
\begin{center}
\includegraphics[width=6.5cm,draft=false]{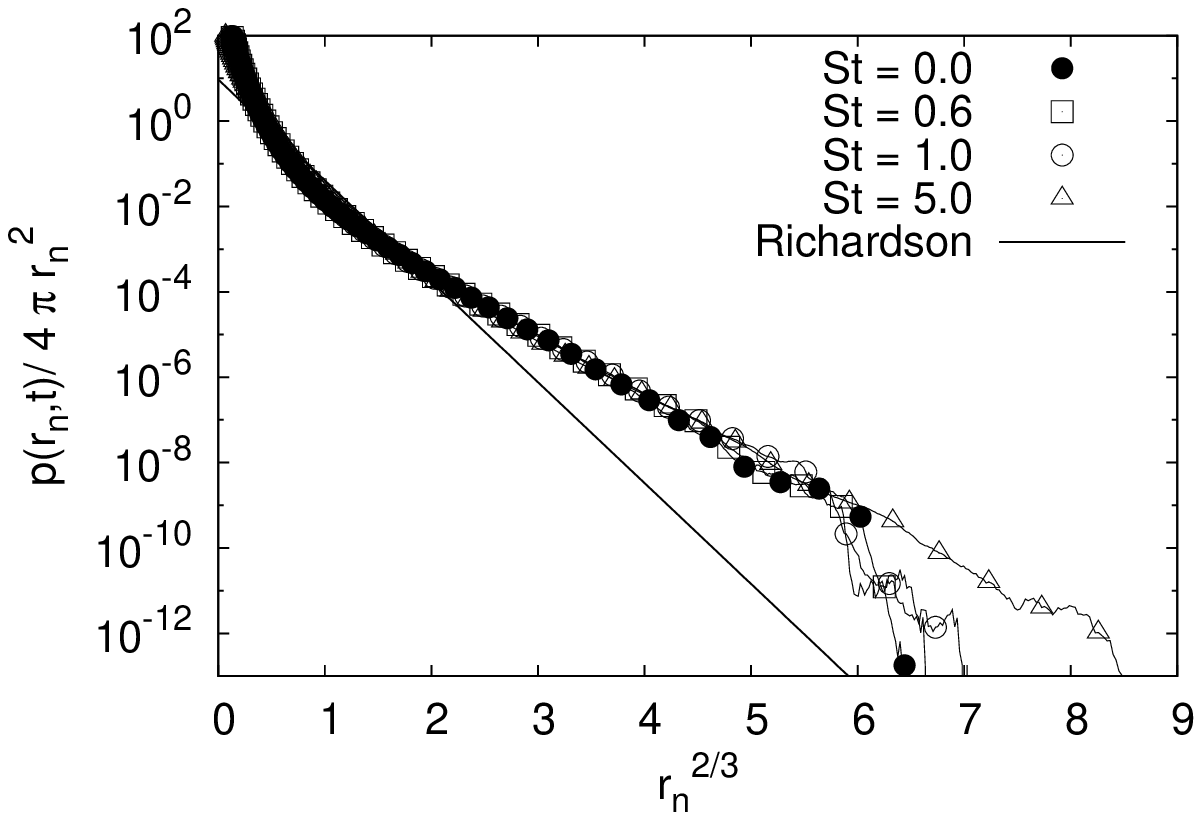}
\includegraphics[width=6.5cm,draft=false]{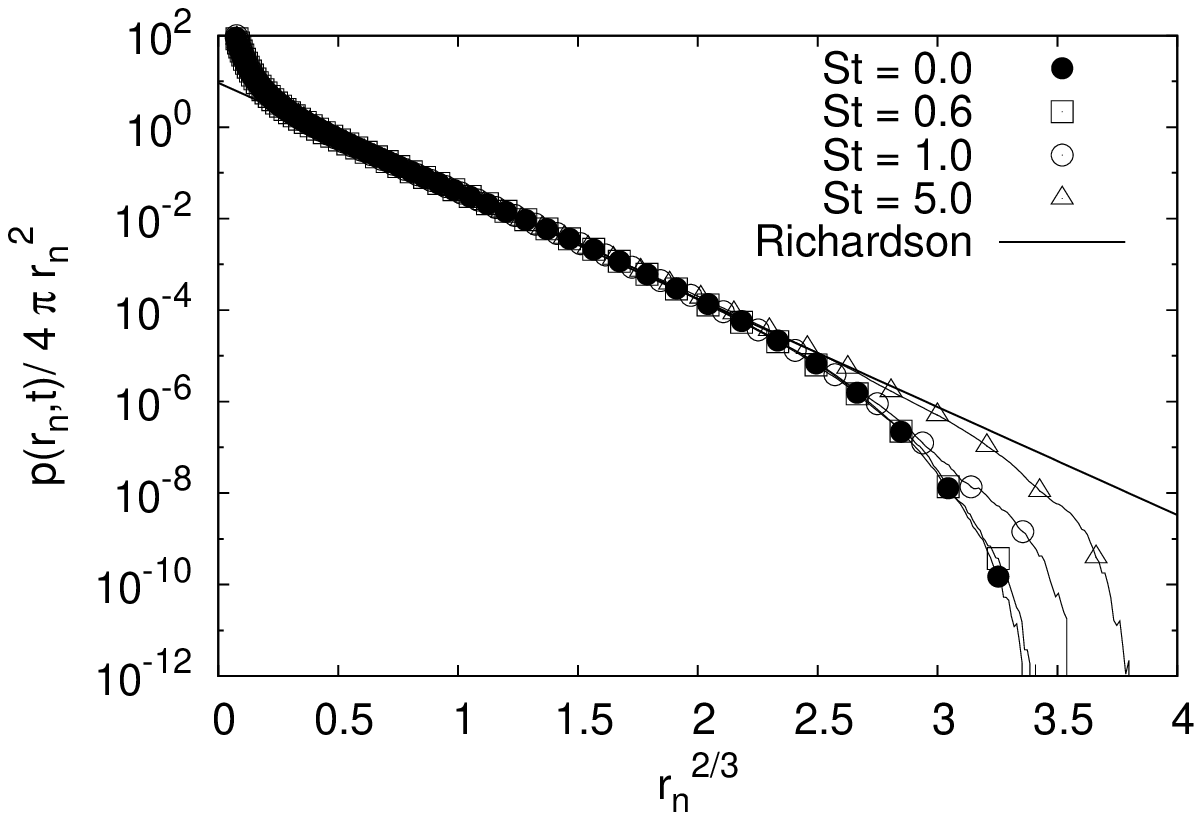}
\caption{Lin-log plot of the pair separation PDFs in rescaled units,
  $r_n = r /\langle r^2(t) \rangle^{1/2}$, at time lags
  $20\tau_{\eta}$ (left panel) and $80\tau_{\eta}$ (right panel), for
  different Stokes numbers.}
 \label{fig_pdf_normalized}
\end{center}
\end{figure}
Differently, we clearly observe that the left tail of the tracer PDF
superposes well with that of the largest Stokes, $St=5$: this is
because the correlation dimension for this high Stokes number is
$D_2(St=5) \simeq d$. At a later time lag, $t \simeq 50 \tau_{\eta}$,
it is very hard to detect a power-law scaling in the left tail, even
with the large database of the present experiment: by this time lag,
most of the pairs have reached larger
separations. Since pair dispersion takes place at
  finite Reynolds numbers, it is clear that asymptotic power-law
  behaviours, see eq.~(\ref{eq:pdfRich}), can be observed in a limited
  range of space and time scales, only. To summarise, the observed
power-law scalings at small separations reproduce the classical
expectations for tracers and heavy pairs reported in \citet{FGV}, and
based on the Richardson's model. To detect intermittency effects, that
we expect to be present due to tracer pairs that separate much less
than the average, different observables are needed. This will be the
object of the exit-time analysis in the last section. \\ Things become
more interesting when the pair separation distribution are plotted in
dimensionless units. In Figure~\ref{fig_pdf_normalized}, we show the
PDFs measured over the whole statistical database, as a function of
the pair distance and also in terms of the normalised relative
separation,
$$ r_n =\frac{r}{\langle r^2(t)\rangle^{1/2}_{St}}.$$ It is important
to notice that the differences as a function of the Stokes number
previously observed are fully reabsorbed once dimensionless quantities
are used. This supports a strong universality for large separations as
a function of $St$, at least up to the values here studied. Should the
PDF data follow the Richardson's prediction, we would see a perfect
rescaling on a stretched exponential curve, for all times and all
separations. It is evident that this is not the case. Moreover, we
stress that the most important departures from the Richardson's
prediction develop on the far right tails, i.e. for intense
fluctuations due to pairs that separate much more than the
average. This fact was already observed using the same dataset by
\citet{SBT12}. Previous numerical and experimental studies were
limited to events with a probability larger than $10^{-6}$ (see e.g.,
\citet{OM00,Bif2005a}, where departures from the Richardson's
prediction could not be unambiguously detected.\\ It must be noticed
that the renormalisation in terms of the separation $r_n$ brings some
extra difficulties in the interpretation of data. Indeed, since the
mean squared separation, $\langle r^2(t)\rangle^{1/2}_{St}$, is
increasing with time, at large times it might well happen that the far
right tails of the PDFs are completely dominated by large-scale
effects, $ r \sim L$. The opposite happens for the events close to the
peak, which can be affected by viscous contributions, $r \sim \eta$,
for small times. In both cases, finite Reynolds-number effects come
into play. As a result, these rescalings do not allow straightforward
conclusions, neither to confirm neither to exclude the alleged
departures from the distribution predicted by Richardson in the
infinite Reynolds limit.\\
\begin{figure}
\begin{center}
\includegraphics[width=6.5cm,draft=false]{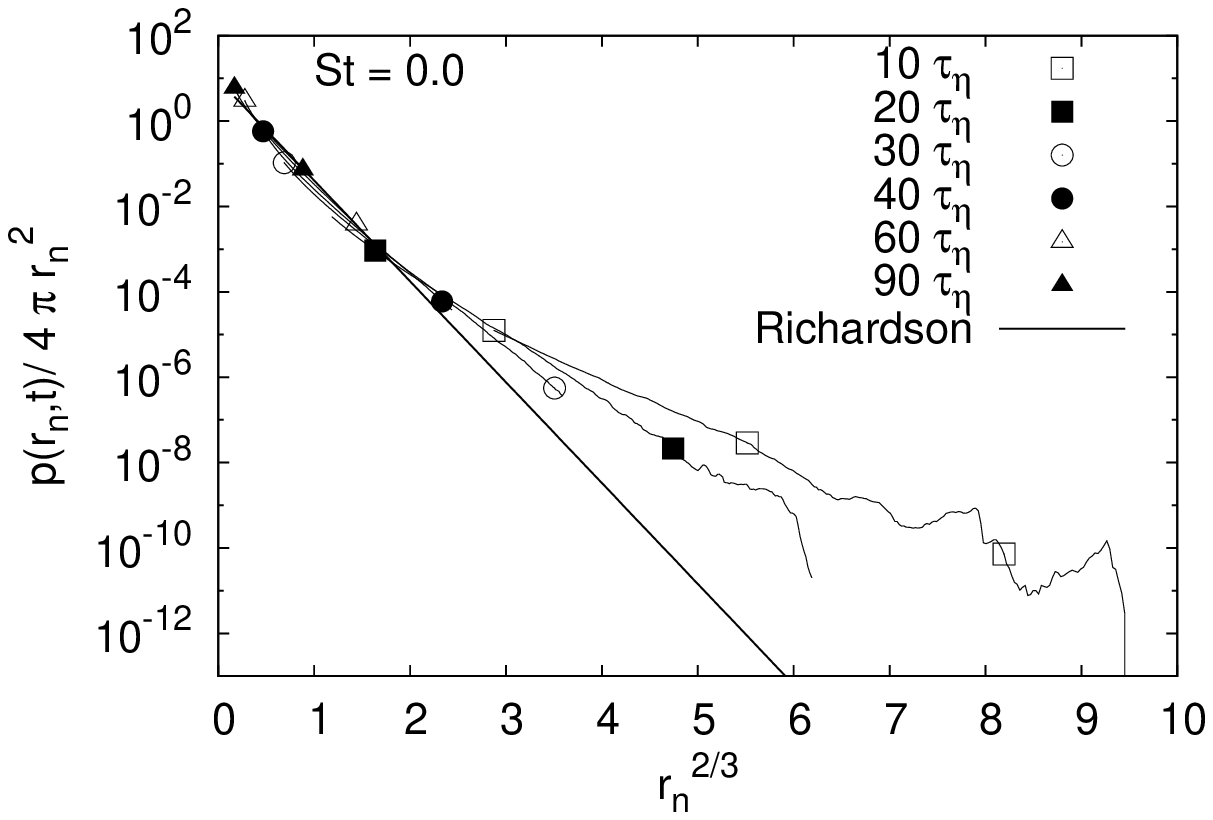}
\includegraphics[width=6.5cm,draft=false]{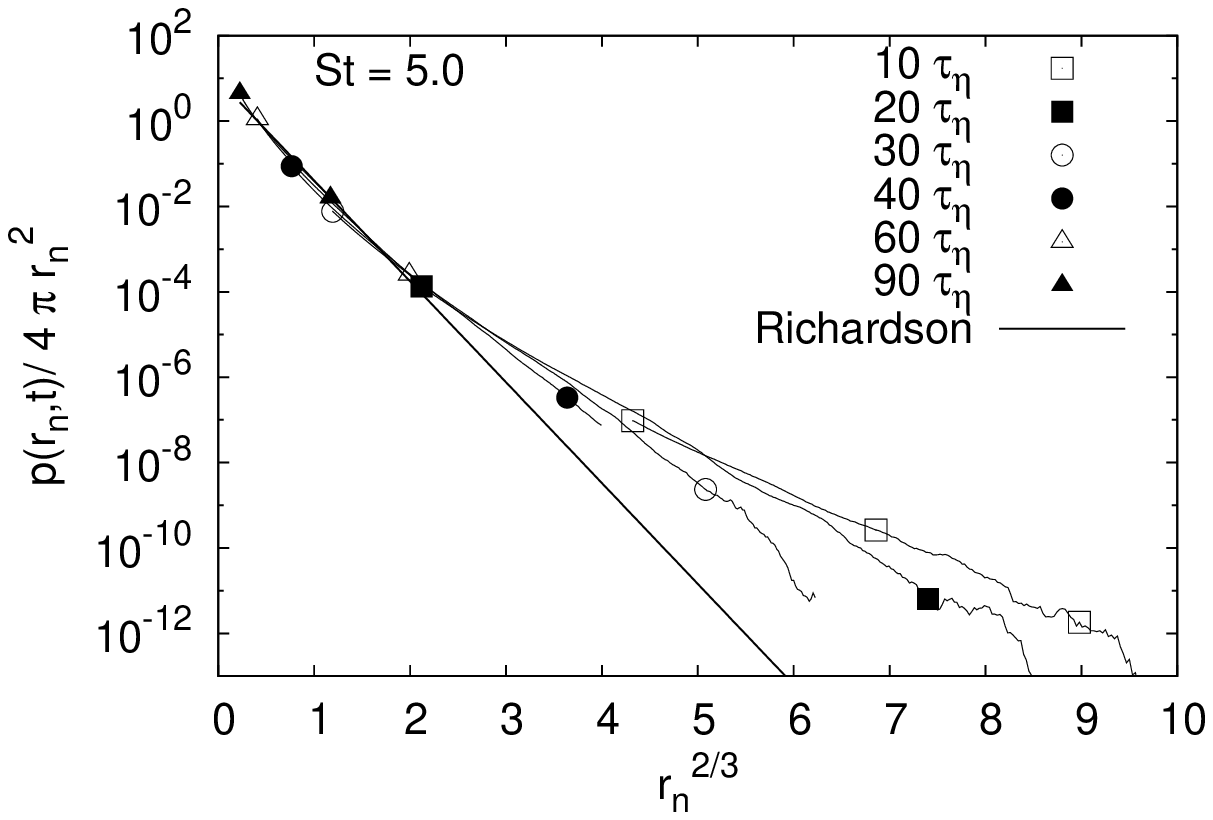}
\caption{Lin-log plot of the rescaled pair separation PDFs at times
    lags $t=(10,20,30,40,60,90)\tau_{\eta}$, selecting the pair
    distances to be in the range $r \in [25, 300] \eta$. Left panel is
    for tracers, while the right panel is for heavy pairs with $St=5$. Symbols are drawn for a subset of points only for clarity.}
  \label{fig_pdf_inertial}
\end{center}
\end{figure}
In Figure~\ref{fig_pdf_inertial} we show the same data of
Figure~\ref{fig_pdf_normalized}, but conditioning the relative
separation to belong to the inertial range of scales, $ 25 \eta < r <
300 \eta$. A more coherent picture now emerges, since we note that (i)
curves belonging to different time lags develop clearly non
overlapping tails; (ii) for times large enough a universal, Stokes
independent, regime seems to develop; (iii) the rapid fall-off of the
left tails of the PDFs for extreme separations disappears. These
observations suggest the possibility to identify inertial range
statistical properties that show a Reynolds-independent departure from the
Richardson's prediction and that can not be attributed to viscous or
large-scale effects.

\subsection{The Multifractal prediction for pair dispersion}
\label{sec:direct-multi}
\begin{figure}
\center
\includegraphics[width=9cm,draft=false]{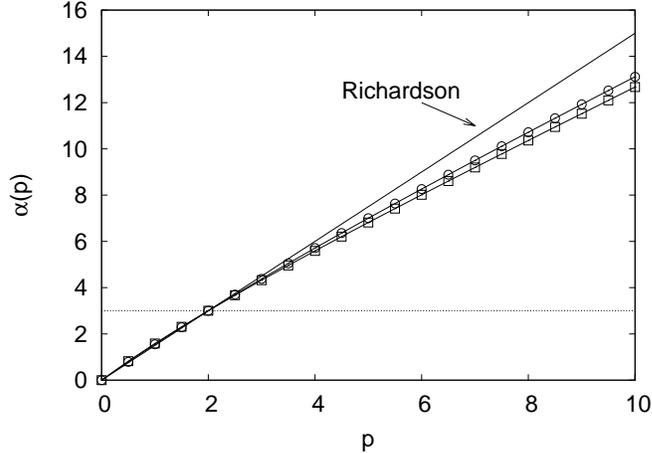}
\caption{Multifractal exponents for pair separation statistics,
  derived from the scaling exponents of the Eulerian longitudinal
  structure functions, $D_L(h)$ ($\Circle$), and from the scaling exponents of Eulerian
  transversal structure functions, $D_T(h)$ ($\square$). The
  continuous line is the dimensional Richardson scaling, $\alpha(p)=
  3p/2$.}
  \label{fig_multi}
\end{figure}

\begin{figure}
\begin{center}
\includegraphics[width=13cm,draft=false]{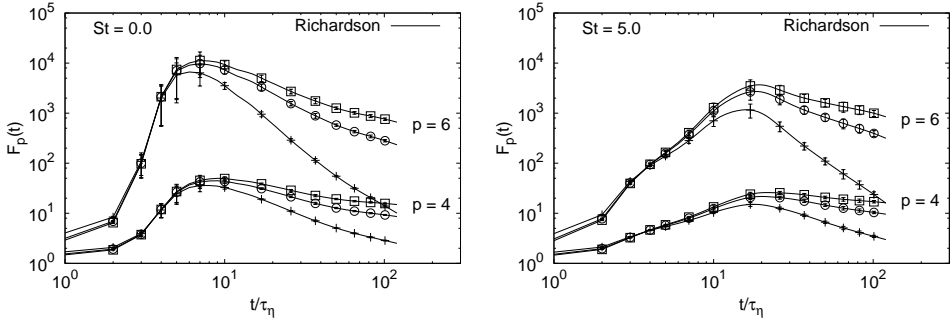}
\caption{The ratio of pair separation moments of order $p=4$ and $p=6$
  to the second order one, as a function of time. For each moment, the
  upper curve is obtained by compensating with the multifractal
  exponent, $\alpha(p)$, obtained from the Eulerian transverse
  structure functions, while the middle curve is obtained by
  compensating with the multifractal exponent from the Eulerian
  longitudinal structure functions (see previous figure and discussion
  in the text). The lower curve is obtained by compensating with the
  Richardson's dimensional scaling $\alpha(p)=3p/2$. (Left) Moments of
  tracer separation; (right) moments of separation of heavy pairs with
  $St=5$. The error bars are given by the root mean
    square of equation~(\ref{eq:ess}) computed on two equal
    sub-ensembles of the whole statistics.}.
  \label{fig_fit}
\end{center}
\end{figure}

\begin{figure}
\begin{center}
\includegraphics[width=13.5cm,draft=false]{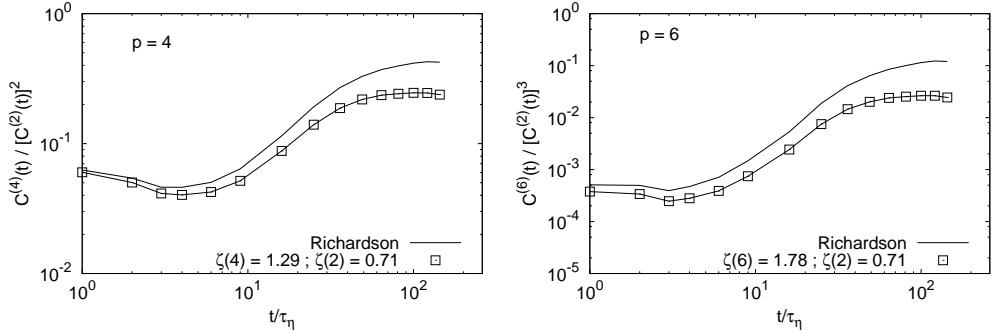}
\caption{The ratio of mixed separation-velocity correlations
  $C^{(p)}(t)$ of order $p=6$ and $p=4$, with respect to a reference
  order, $C^{(2)}(t)$, vs time. The moments are compensated with the
  dimensional prediction of the theory of Richardson, and with the
  longitudinal intermittent scaling exponents.}
  \label{fig_grisha}
\end{center}
\end{figure}

The presence of inertial-range deviations from the pure self-similar
behaviour of pair separation statistics predicted by the Richardson's
approach should not be surprising, although never observed. In the
$3d$ direct energy cascade regime, anomalous scaling is measured both
in the statistics of Eulerian longitudinal and transverse velocity
increments \citep{frisch,sreeni,BifPhysD}, and in the statistics of
Lagrangian velocity increments along single particle trajectories
\citep{Pinton,BifPRL04,XuPRL,ICTRPRL,BCT2011}. A simple argument
predicts the presence of intermittent corrections in the high-order
moments of the relative particles separation, $\langle r^p \rangle$
\citep{N89,BCCV99}. The starting point is the exact relation for the
moment of order $p$ of the pair separation,
\begin{equation}
\frac{d}{dt} \langle r^p\rangle = p \langle r^{p-1} (\delta_r
u)\rangle\,,
\label{eq:rp}
\end{equation}
where $\delta_r u$ is the velocity increment measured along the tracer
pair trajectories, separated by a distance $r$. Here, for simplicity,
we have neglected the tensorial structure and the time dependency of
$r$ and $\delta_r u$ is understood. Let us suppose
  that the above correlation can be estimated with fully Eulerian
  quantities, then the multifractal approach \cite{frisch} could be
  employed to obtain:
\begin{equation}
\label{eq:multi}
\langle r^{p-1} (\delta_{r} u)\rangle \propto \int dh\, r^{3-D(h)}
 r^{p-1} r^{h}\,,
\end{equation}
where $P(h) \propto r^{3-D(h)}$ is the probability to observe an
Eulerian velocity fluctuation at the scale $r$ with a local scaling
exponent $h$, $\delta_r u \sim r^h$, as a function of the spectrum of
fractal dimension, $D(h)$. In order to relate the above Eulerian
estimate with the Lagrangian one in (\ref{eq:rp}), one can adopt the
dimensional {\it bridge} relation supposing that Lagrangian velocity
fluctuations at separation $r(t)$ are connected to Eulerian spatial
fluctuations at scale $r$ with $ t \sim r/\delta_r u \sim r^{1-h}$.
This amounts to say that one can use the same fractal dimensions
$D(h)$ for the Lagrangian and Eulerian velocity statistics. A
quantitative support for this hypothesis has been made by a validation
for one particle quantities against numerical and experimental results
in \cite{ICTRPRL}. The same argument applied to two-particle
quantities allows to rewrite (\ref{eq:multi}) as a function of the
time lag, $t$, along the particle separations,
\begin{equation}
\label{eq:multi_time}
 \langle r^{p-1} (\delta_{r} u)\rangle \propto \int d h\,
 t^{\frac{2-D(h)+p+h}{1-h}}\,.
\end{equation}
After time integration, a saddle point approximation can be used in
the limit $t \rightarrow 0$, when the smallest exponent dominates the
integral \citep{BCCV99}:
\begin{equation}
\label{eq:multi_espo}
\langle r^p(t)\rangle \propto t^{\alpha(p)},\qquad \alpha(p) = \min_h{\frac{(3-D(h)+p)}{(1-h)}}\,.
\end{equation}
This is the multifractal theory for tracer pair separation statistics
in the presence of an Eulerian velocity field \citep{BCCV99}, whose
multiscale statistics is described by the set of fractal dimension
$D(h)$. To our knowledge, such a prediction has never been tested,
either on experimental or on numerical data, except for the
validation against stochastic velocity fields reported in
\citet{BCCV99}. Note that the multifractal spectrum is in general
non-linear in the order $p$; since by the $4/5$-law, the exponent of
the third order longitudinal Eulerian structure function must be
unity, we must also have that $\alpha(2)\equiv 3$
\citep{N89,BCCV99}. In Figure~\ref{fig_multi}, we compare the
multifractal spectrum, $\alpha(p)$, obtained by using two possible
forms for $D(h)$ (extracted, respectively, from the longitudinal
Eulerian structure functions, $D_L(h)$, and from the transverse ones,
$D_T(h)$) in statistically homogeneous and isotropic $3d$
turbulence. Such small uncertainty is to be considered as our
prediction error on the shape of the function $D(h)$, which cannot be
deduced by first principles. A more detailed discussion about this
point can be found in \citet{benzi2010}.\\
To measure the scaling behaviours, we use the procedure known as
Extended Self Similarity (ESS) \cite{ess}, which amounts to study the
relative scaling of moments with respect to a reference one, whose
exponent is constrained by an exact relation. In Figure~\ref{fig_fit},
the $p-$th order moment of particle separation is compensated with the
second order one:
\begin{equation}
F_p(t) = \frac{\langle r^p(t) \rangle}{\langle r^2(t)
  \rangle^{\frac{\alpha(p)}{3}}}\,,
\label{eq:ess}
\end{equation}
both for tracer pairs and for heavy pairs at $St=5$.
In the inertial range where the prediction of
  equation (\ref{eq:multi_espo}) is expected to be valid, we should
  observe a plateau. It is evident that the multifractal prediction
  works better than the dimensional one, suggesting that the approach
  goes in the right direction. However, because the plateau is very
  narrow it is necessary to wait for data at higher Reynolds before
  making any firm conclusion.
The claim is that the observed departure from the multifractal
prediction in the tracers statistics is due to contamination induced
by viscous effects, which we have seen to be very strong for the
$St=0$ case. The situation becomes more interesting for heavy
particles at $St=5$ for which viscous effects have a smaller
impact. Here the multifractal scaling gives a larger plateau for the
$p=4$ moment, and shows the beginning of a plateau for the $p=6$
moment. We thus have an indication of an inertial-range intermittent
effect for particle pair separation statistics in homogeneous and
isotropic $3d$ turbulence. \\ This is the main result of this
paper. \\
On the basis of the multifractal formalism, it is straightforward to
conclude that there exist correlation functions that should be {\it
  statistically preserved} along pair trajectories, for scales well
within the inertial range \citep{FGV,FalkPRL2013}. Indeed, by again
applying the saddle-point estimate to the inertial range scaling of
the mixed separation-velocity moments, according to multifractal model
we get that
\begin{equation}
\label{eq:multi2}
C^{(p)}(t)\, = \, \langle r(t)^{-\zeta(p)} (\delta_{r(t)} u)^p\rangle \sim const.\,
\end{equation}
if the scaling exponents of the particle separation compensates that
of the Eulerian velocity moment: $\zeta(p) \equiv min_h(ph +3
-D(h))$. In Figure~\ref{fig_grisha}, we measure the mixed
separation-velocity correlations of order $p=4$ and $p=6$ in ESS, that
is with reference to the moment of order two. The scaling
obtained by using the $\zeta(p)$ exponents is the one that works
better for inertial range time lags, i.e. $t > 20 \tau_{\eta}$.

\section{Exit-time statistics for tracers}
\label{sec:exit}
\begin{figure}
\begin{center}
 \includegraphics[width=13cm,draft=false]{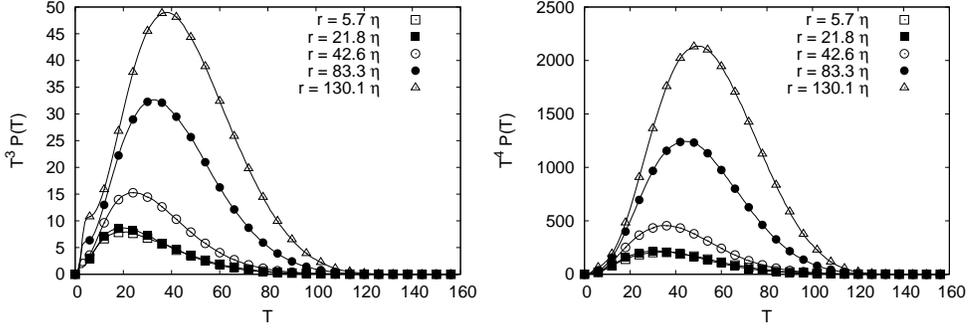}
 \caption{Products $T_{\rho}^p P(T_{\rho})$ with $p=3$ (left) and
   $p=4$ (right) for $\rho = 1.25$, testing the statistical convergence.}
  \label{fig:T3T4p}
\end{center}
\end{figure}

\begin{figure}
 \begin{center}
 \includegraphics[width=9cm,draft=false]{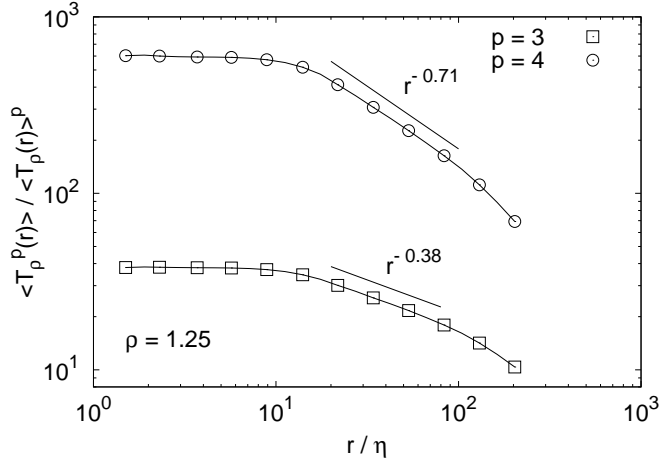}
 \caption{The ratio of positive exit-time moments of order $p=3$ and
   $p=4$ to $\langle T_\rho(r)\rangle^p$, as a function of thresholds
   for $\rho = 1.25$. For different values of $\rho$ we observe the
   same behaviours.}
  \label{fig:ess_exit}
 \end{center}
\end{figure}

Another interesting question concerns the statistical properties of
{\it weak} separation events, i.e. the left tail of the probability
density function. In order to assess the importance of these events,
one cannot resort to negative moments of the separation statistics,
because these are ill defined. An alternative approach which overcomes
this difficulty is to use {\it inverse} statistics, see \cite{Jensen_99},
i.e., exit-times. Exit-times also allow for a clearer separation of
scales as stressed by \citet{BofSok2002_3d}. The idea consists in
fixing a set of thresholds, $r_n=\rho^n r_0$ with $n = 1, 2, 3,...$
and $\rho > 1$, and in calculating the probability density function of
the time $T(r_n)$ needed for the pair separation to change from $r_n$
to $r_{n+1}$. Formally this corresponds to calculating the first
passage time. The advantage of this approach is that all pairs are
sampled when they belong to eddies of similar size,
  between $r_n$ and $r_{n+1}$. This limits the effect of pairs that at
  a given time lag, since they have separated very fast or very slow,
  might be at very different separation scales. In other words, in the
  exit-time statistics the {\it contamination} from viscous and
  large-scale cut-offs should be less important.  
\\ For particle pairs with initial condition $P(r,t=0)=\rho^2\delta(r-r_n/
\rho)/4\pi r_n^2$, a perfectly reflecting boundary condition at $r =
0$ and an absorbing boundary condition at $r = r_n$, the PDF of exit
time, $P(T)$, is given by:
\begin{equation}
\label{eq:pdfexit}
{\cal P}_{\rho, r_n}(T) = -\frac{d}{dt} \int_{|{\bf r}| < r_n} P({\bf r},t) d{\bf r}\,.
\end{equation} 
Using the Richardson's distribution of eq. (\ref{eq:pdfRich}) for
$P({\bf r},t)$, we get
\begin{equation}
\label{eq:pdfexit1}
{\cal P}_{\rho, r_n}(T) = -4 \pi k_0 \epsilon^{1/3} r_n^{10/3}
\partial_r P |_{r=r_n}\,.
\end{equation} 
An asymptotic form of exit-time PDF behaves according to the
following expression,
\begin{equation}
 \label{eq:pdfexit2}
{\cal P}_{\rho, r_n}(T) \simeq  exp \left[-\kappa\, \frac{\rho^{2/3}-1}{\rho^{2/3}}\, \frac{T}{\langle T_\rho(r_n)\rangle}\right]\,,
\end{equation} 
where $\kappa \simeq 2.72$ is a dimensionless constant, for details see
\citet{Bif2005a}, and $\langle T_\rho(r_n)\rangle$ is the mean
exit-time. Note that eq.~(\ref{eq:pdfexit2}) contains only
dimensionless parameters and it is thus a universal result. We note
that while positive moments $\langle T_{\rho}^p(r_n) \rangle$
preferably sample pairs that separate slowly, negative moments,
$\langle \left[1/T_{\rho}(r_n)\right]^p \rangle$, are dominated by
pairs that separate fast, \cite{BofSok2002_3d}. From
eq.~(\ref{eq:pdfexit2}), a prediction can be obtained for the mean
exit-time \citep{BofSok2002_3d},
\begin{equation}
\label{eq:meanexit}
\langle T_{\rho}(r) \rangle = \frac{1}{2\,k_0\,\epsilon^{1/3}}\frac{(\rho^{2/3}-1)}{\rho^{2/3}} \, r^{2/3}\,,
\end{equation} 
from which it follows that according to the Richardson self-similar
behaviour, we expect
\begin{equation}
\label{eq:exitdimen}
\langle T_{\rho}^p(r) \rangle \propto r^{2p/3}\,.
\end{equation}
\begin{figure}
\begin{center}
 \includegraphics[width=12.5cm,draft=false]{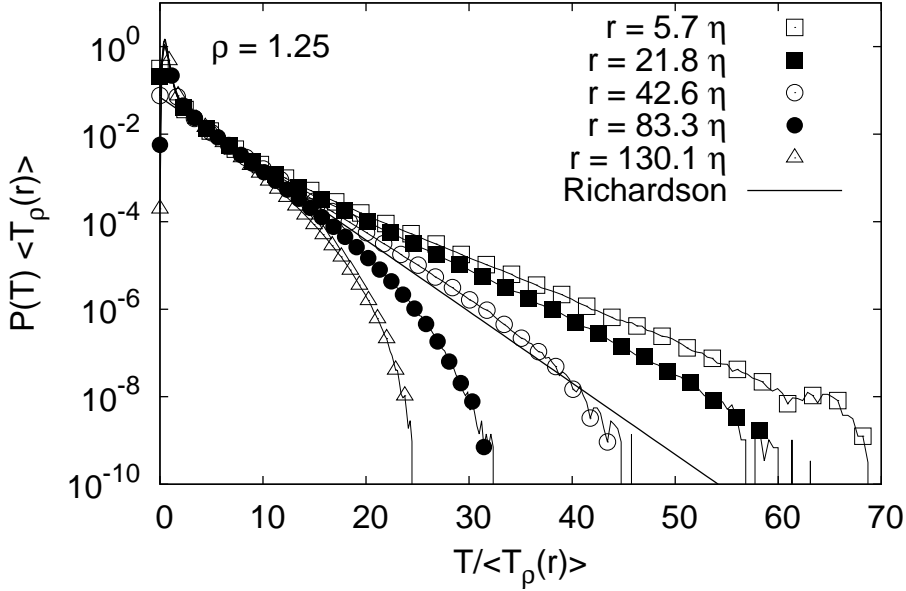}
 \caption{The exit-time PDFs for tracers pairs are shown, together
   with the Richardson's asymptotic form of the exit-time
   distribution. The growth factors for spatial thresholds is
   $\rho=1.25$. The continuous straight line is the Richardson's
   prediction.}
  \label{fig:10}
\end{center}
\end{figure}
In Figure~\ref{fig:10}, we plot the exit-time PDFs for the tracer
pairs, calculated for $\rho=1.25$. First, let us notice that the
super-exponential decay observed for the large distance case is
probably due to a systematic bias induced by the fact that we have a
finite length in the trajectories and therefore very long exit times
cannot be measured. Second, we observe that the curves do not overlap
meaning that the PDFs are not self-similar. Concerning the statistical
accuracy, this result is a clear improvement of the one reported in
\citet{Bif2005a}: now deviations from the Richardson's prediction are
evident because of the huge statistics achieved in the present
numerical experiment. Even though the asymptotic distribution is still
given by an exponential decay, as it should be expected from rare
events following a Poissonian process, the whole PDF shape cannot be
superposed using only the mean exit-time as a normalising
factor.\\ Concerning the positive moments of the exit times, we use
relative scaling properties to test a breaking of the self-similar
properties. In Figure \ref{fig:T3T4p} we verify that we have the
statistical convergence needed to measure moments of order $\langle
T^3_\rho(r)\rangle$ and $\langle T^4_\rho(r)\rangle$. In
Figure~\ref{fig:ess_exit}, we show the ratio of $T_{\rho}(r)$ moments
of order $p=3,4$ with respect to $\langle T_\rho(r)\rangle^p$, for
$\rho = 1.25$.  For separations within the inertial range $r/\eta >
O(10)$ the breaking of self-similarity is evident.
It is difficult to conclude if these are
true Reynolds independent corrections, and if they are affected by the
finite length of the particle trajectories. More data will be needed
to get a more quantitative understanding of this effect. Furthermore,
we recall that there is not a multifractal prediction for the
behaviour of the positive exit-time moments of relative dispersion,
since these would be associated to the negative moments of the
Eulerian velocity increments \citep{Jensen_99}.
\section{Conclusions}

  In this work we have numerically studied the
  relative dispersion statistics of tracer and heavy particles emitted
  from point-like sources in a homogeneous and isotropic turbulent
  flow, at resolution $1024^3$. When particles are injected at
  separations of the order of the viscous scale, the fluctuations of
  the local stretching rate have a huge impact, influencing the mean
  square separation, up to time lags of the order of the integral
  time-scale. Such a dramatic effect has hindered the possibility of
  studying inertial range quantities in Lagrangian experiments or
  numerical simulations up to now. In this study, the large
  statistical database, the use of conditional statistics and the
  information obtained by comparing tracers and inertial particles
  evolution enabled us to highlight with a great precision, and for
  the first time, deviations in the pair separation distribution from
  the self-similar behaviour predicted by Richardson. Such deviations
  are manifest in the statistical behaviour of the right tail of the
  separation PDF, and are due to tracer pairs that separate much
  faster than the average. We use similar measurements for heavy
  particles at moderate inertia, which filter out fluctuations of the
  viscous scales, to give a higher confidence that the corrections
  observed for tracers are pure inertial range effect. Moreover, by
  conditioning the relative separation to belong to the inertial range
  of scales, a universal behaviour develops, which is Stokes
  independent and is clearly different from the Richardson's
  prediction. The behaviour of the conditioned PDFs support the idea
  that finite Reynolds-number effects, even if present, are
  sub-dominant. Furthermore, the numerical results indicate that
  tracer dispersion is intermittent, with deviations from a
  self-similar scaling visible already in the low order moments
  behaviour. The observed intermittent corrections to the Richardson's
  prediction are qualitatively consistent with a multifractal
  prediction for the scaling behaviour of relative separation moments
  of tracer pairs, although in a narrow scaling region. Numerical and
  experimental results at higher Reynolds number are requested to
  further support these findings. By measuring the exit-time
  statistics, we provide also an evidence of the non self-similar 
  character of slow pair dispersion
  events. The statistics of the shapes of the puffs along the temporal
  evolution is also a crucial point that worths to be studied,
  involving more information about the multi-particle separation
  connected to geometry and shapes \cite{Cher1999,Bif2005b,Xu2011}

\acknowledgments We thank J\'er\'emie Bec and Gregory Eyink for useful
discussions. We acknowledge financial support from EU-COST Action
MP0806 ``Particles in Turbulence''. This work is part of the research
program of the Foundation for Fundamental Research on Matter (FOM),
which is part of the Netherlands Organisation for Scientific Research
(NWO). Numerical simulations were performed within the HPC Iscra Class
A projects ``Point'' and ``Flag'' at CINECA (Italy). L.B. acknowledge
partial funding from the European Research Council under the European
Community's Seventh Framework Programme, ERC Grant Agreement
N. 339032.

\bibliographystyle{jfm}

\end{document}